\documentclass[]{JHEP3}
\usepackage{epsfig}

\def\be{\begin{equation}}
\def\ee{\end{equation}}
\def\bea{\begin{array}}
\def\eea{\end{array}}
\def\beqa{\begin{eqnarray}}
\def\eeqa{\end{eqnarray}}
\def\beqas{\begin{eqnarray*}}
\def\eeqas{\end{eqnarray*}}

\def\bp{\begin{picture}}
\def\ep{\end{picture}}
\def\bc{\begin{center}}
\def\ec{\end{center}}
\def\bfig{\begin{figure}}
\def\efig{\end{figure}}

\def\bit{\begin{itemize}}
\def\eit{\end{itemize}}
\def\nn{\nonumber}
\def\f{\frac}

\def\[{\left[}
\def\]{\right]}
\def\({\left(}
\def\){\right)}

\def\..{\left.}
\def\.{\right.}
\def\tl{\widetilde}
\def\ra{\rightarrow}

\def\tm{\times}

\def\ep{\epsilon}

\def\de{\delta}

\def\pr{\prime}

\title{ Supersymmetry Breaking Scalar Masses and Trilinear Soft Terms
in Generalized Minimal Supergravity}
\author{Csaba Balazs$^1$, Tianjun Li$^{2,3}$, Dimitri V.
Nanopoulos$^{2,4,5}$,
Fei Wang$^1$ \\
$^1$ School of Physics, Monash University, Melbourne Victoria 3800,
Australia\\
$^2$ George P. and Cynthia W. Mitchell Institute for
Fundamental Physics, Texas A$\&$M University, College Station, TX
77843, USA \\
$^3$ Key Laboratory of Frontiers in Theoretical Physics,
Institute of Theoretical Physics, Chinese Academy of Sciences,
Beijing 100190, P. R. China \\
$^4$ Astroparticle Physics Group,
Houston Advanced Research Center (HARC),
Mitchell Campus, Woodlands, TX 77381, USA \\
$^5$ Academy of Athens, Division of Natural Sciences,
28 Panepistimiou Avenue, Athens 10679, Greece
}
\abstract{
In the generalized minimal supergravity (GmSUGRA)
scenario, we systematically study the supersymmetry breaking
scalar masses, Standard Model fermion Yukawa coupling terms,
and trilinear soft terms in $SU(5)$ models with
the Higgs fields in the ${\bf 24}$ and ${\bf 75}$
representations, and in $SO(10)$ models where
the gauge symmetry is broken down to the Pati-Salam
$SU(4)_C\times SU(2)_L \times SU(2)_R$ gauge symmetry,
$SU(3)_C\times SU(2)_L \times SU(2)_R \times U(1)_{B-L}$
gauge symmetry, George-Glashow $SU(5)\times U(1)'$
gauge symmetry, flipped $SU(5)\times U(1)_X$ gauge symmetry,
and $SU(3)_C\times SU(2)_L \times U(1)_1 \times U(1)_2$
gauge symmetry. Most importantly, we for the first time
consider the scalar and
gaugino mass relations, which can be preserved from
the unification scale to the electroweak scale
under one-loop renormalization group equation running, in the $SU(5)$
models, the Pati-Salam models and flipped $SU(5)\times U(1)_X$
models arising from $SO(10)$ models.
With such interesting relations, we may distinguish the
minimal supergravity (mSUGRA) and GmSUGRA scenarios
if the supersymmetric particle spectrum can be
measured at the LHC and ILC. Thus, it provides us with
another important window of opportunity at the Planck scale.
}
\preprint{ACT-09-10, MIFPA-10-26}

\begin{document}
\maketitle \indent
\newpage

\section{Introduction}

Supersymmetry naturally solves the
gauge hierarchy problem of the Standard Model (SM). The unification
of the three gauge couplings $SU(3)_C, SU(2)_L$ and $U(1)_Y$ in the
supersymmetric Standard Model at about $2\tm 10^{16}$
GeV~\cite{Ellis:1990zq} strongly suggests the existence of Grand
Unified Theories (GUTs). In addition,
supersymmetric GUTs such as $SU(5)$~\cite{Georgi:1974sy} or
$SO(10)$~\cite{so10} models give us deep insights into the
other SM problems such as the emergence of the fundamental forces,
the assignments and quantization of their charges,
the fermion masses and mixings, and beyond.
Although supersymmetric GUTs are
attractive it is challenging to test them
at the Large Hadron Collider (LHC), the future
International Linear Collider (ILC), and other experiments.

In traditional supersymmetric SMs
supersymmetry is broken in the hidden sector and
the supersymmetry breaking effects can be mediated
to the observable sector via gravity~\cite{mSUGRA},
gauge interactions~\cite{Ellis:1984bm, gaugemediation},
or super-Weyl
anomaly~\cite{anomalymediation, UVI-AMSB, D-AMSB}, or other mechanisms.
However, the relations between the supersymmetric particle
(sparticle) spectra and the fundamental theories can be very
complicated and model dependent. An important observation is
that compared to the
supersymmetry breaking soft masses
of squarks and sleptons (scalar masses),
gaugino masses have a simpler form and are
less model dependent~\cite{Ellis:1985jn, Choi:2007ka}.
In the minimal supergravity (mSUGRA)
scenario~\cite{mSUGRA} supersymmetry breaking is mediated
by gravity and gauge couplings and gaugino masses are unified
at the GUT scale. Thus, a relation holds between the
the gauge couplings and the gaugino masses at the GUT scale
$M_{\rm GUT}$:
\begin{eqnarray}
{{1}\over {\alpha_3}} ~=~ {{1}\over {\alpha_2}}
~=~ {{1}\over {\alpha_1}} ~,~\,
\label{mSUGRA-C}
\end{eqnarray}
\begin{eqnarray}
{{M_3}\over {\alpha_3}} ~=~ {{M_2}\over {\alpha_2}}
~=~ {{M_1}\over {\alpha_1}} ~,~\,
\label{mSUGRA}
\end{eqnarray}
where $\alpha_3$, $\alpha_2$, and $\alpha_1\equiv 5\alpha_{Y}/3$
($M_3$, $M_2$, and $M_1$) are gauge couplings (gaugino masses)
for the $SU(3)_C$, $SU(2)_L$, and $U(1)_Y$ gauge symmetries,
respectively.
Because $M_i/\alpha_i$ are
constant under renormalization group evolution,
the gaugino mass relation
in Eq.~(\ref{mSUGRA}) is valid from the GUT scale to the
electroweak scale at one loop.
Two-loop renormalization group effects on gaugino masses
are very small, thus, we can test this gaugino mass relation
at the LHC and ILC where the gaugino masses can be
measured~\cite{Cho:2007qv, Barger:1999tn}.
Recently, considering GUTs with high-dimensional
operators~\cite{Ellis:1984bm, Ellis:1985jn, Hill:1983xh,
Shafi:1983gz, Drees:1985bx,
Anderson:1999uia, Chamoun:2001in, Chakrabortty:2008zk, Martin:2009ad,
Bhattacharya:2009wv, Feldman:2009zc, Chamoun:2009nd}
and F-theory GUTs with $U(1)$ fluxes~\cite{Vafa:1996xn,
Donagi:2008ca, Beasley:2008dc, Beasley:2008kw, Donagi:2008kj,
Font:2008id, Jiang:2009zza, Blumenhagen:2008aw, Jiang:2009za,
Li:2009cy, Leontaris:2009wi, Li:2010mr},
two of us (TL and DN) proposed the generalized
 mSUGRA (GmSUGRA) scenario, and studied
the generic gaugino mass relations and defined
their indices~\cite{Li:2010xr}.
The gaugino mass relations and their indices
have also been studied for
general gauge and anomaly mediated supersymmetry breaking in
GUTs with vector-like particles~\cite{Li:2010hi}.

In this paper, we
consider the supersymmetry breaking
scalar masses and trilinear soft terms in the GmSUGRA.
We briefly review GUTs and consider the general
gravity mediated supersymmetry breaking. With the
high-dimensional operators including the GUT Higgs fields,
we systematically calculate the supersymmetry breaking scalar masses,
SM fermion Yukawa coupling terms,
and trilinear soft terms in $SU(5)$ models with
GUT Higgs fields in the ${\bf 24}$ and ${\bf 75}$
representations, and in $SO(10)$ models where
the gauge symmetry is broken down to the Pati-Salam
$SU(4)_C\times SU(2)_L \times SU(2)_R$ gauge symmetry,
$SU(3)_C\times SU(2)_L \times SU(2)_R \times U(1)_{B-L}$
gauge symmetry, George-Glashow $SU(5)\times U(1)'$
gauge symmetry, flipped $SU(5)\times U(1)_X$ gauge
symmetry~\cite{smbarr, dimitri, AEHN-0},
and $SU(3)_C\times SU(2)_L \times U(1)_1 \times U(1)_2$
gauge symmetry. We examine the scalar and
gaugino mass relations, which are valid from
the GUT scale to the electroweak scale
under one-loop renormalization group running, in the $SU(5)$
models, the
Pati-Salam models and flipped $SU(5)\times U(1)_X$
models arising from the $SO(10)$ model.
With these relations, we may distinguish the
mSUGRA and GmSUGRA scenarios if the supersymmetric
particle spectrum can be measured at the LHC and ILC.

This paper is organized as follows. In Section~2,
we briefly review four-dimensional GUTs.
In Section~3, we explain the general gravity mediated
supersymmetry breaking. In Section~4,
we discuss the scalar masses, the SM fermion Yukawa
coupling terms, and trilinear soft terms
in the $SU(5)$ model. For models arising from $SO(10)$,
we derive the scalar
masses in Section~5, and the SM fermion Yukawa
coupling terms and trilinear soft terms in Section~6.
In Section~\ref{sec-4} we consider
the scalar and gaugino mass relations.
Section~\ref{sec-5} contains our conclusions.

\section{ Brief Review of Grand Unified Theories}
\label{sect-0}

In this Section we explain our conventions.
In supersymmetric SMs,
we denote the left-handed quark doublets, right-handed
up-type quarks, right-handed down-type quarks,
left-handed lepton doublets, right-handed neutrinos
and right-handed charged leptons as $Q_i$, $U^c_i$, $D^c_i$,
$L_i$, $N^c_i$, and $E^c_i$, respectively. Also, we denote
one pair of Higgs doublets as $H_u$ and $H_d$, which give masses
to the up-type quarks/neutrinos and the down-type quarks/charged
leptons, respectively.

First, we briefly review the $SU(5)$ model.
We define the $U(1)_{Y}$ hypercharge generator in $SU(5)$ as follows
\begin{eqnarray}
T_{\rm U(1)_{Y}}={\rm diag} \left(-{1\over 3}, -{1\over 3}, -{1\over 3},
{1\over 2}, {1\over 2} \right)~.~\,
\label{u1y}
\end{eqnarray}
Under $SU(3)_C\times SU(2)_L \times U(1)_Y$ gauge symmetry,
the $SU(5)$ representations
are decomposed as follows
\begin{eqnarray}
\mathbf{5} &=& \mathbf{(3, 1, -1/3)} \oplus \mathbf{(1, 2, 1/2)}~,~ \\
\mathbf{\overline{5}} &=&
\mathbf{(\overline{3}, 1, 1/3)} \oplus \mathbf{(1, 2, -1/2)}~,~ \\
\mathbf{10} &=& \mathbf{(3, 2, 1/6)} \oplus \mathbf{({\overline{3}}, 1,
-2/3)}
\oplus \mathbf{(1, 1, 1)}~,~ \\
\mathbf{\overline{10}} &=& \mathbf{(\overline{3}, 2, -1/6)}
\oplus \mathbf{(3, 1, 2/3)}
\oplus \mathbf{(1, 1, -1)}~,~ \\
\mathbf{24} &=& \mathbf{(8, 1, 0)} \oplus \mathbf{(1, 3, 0)}
\oplus \mathbf{(1, 1, 0)} \oplus \mathbf{(3, 2, -5/6)} \oplus
\mathbf{(\overline{3}, 2, 5/6)}~.~\,
\end{eqnarray}
There are three families of the SM fermions
whose quantum numbers under $SU(5)$ are
\begin{eqnarray}
F'_i=\mathbf{10},~ {\overline f}'_i={\mathbf{\bar 5}},~
N^c_i={\mathbf{1}}~,~
\label{SU(5)-smfermions}
\end{eqnarray}
where $i=1, 2, 3$ for three families.
The SM particle assignments in $F'_i$ and ${\bar f}'_i$ are
\begin{eqnarray}
F'_i=(Q_i, U^c_i, E^c_i)~,~{\overline f}'_i=(D^c_i, L_i)~.~
\label{SU(5)-smparticles}
\end{eqnarray}

To break the $SU(5)$ gauge symmetry and electroweak gauge symmetry,
we introduce the adjoint Higgs field and one pair
of Higgs fields whose quantum numbers under $SU(5)$ are
\begin{eqnarray}
\Phi'~=~ {\mathbf{24}}~,~~~
h'~=~{\mathbf{5}}~,~~~{\overline h}'~=~{\mathbf{\bar {5}}}~,~\,
\label{SU(5)-1-Higgse}
\end{eqnarray}
where $h'$ and ${\overline h}'$ contain the Higgs doublets
$H_u$ and $H_d$, respectively.

Second, we briefly review the flipped
$SU(5)\times U(1)_{X}$ model~\cite{smbarr, dimitri, AEHN-0}.
The gauge group $SU(5)\times U(1)_{X}$ can be embedded
into $SO(10)$.
We define the generator $U(1)_{Y'}$ in $SU(5)$ as
\begin{eqnarray}
T_{\rm U(1)_{Y'}}={\rm diag} \left(-{1\over 3}, -{1\over 3}, -{1\over 3},
{1\over 2}, {1\over 2} \right).
\label{u1yp}
\end{eqnarray}
The hypercharge is given by
\begin{eqnarray}
Q_{Y} = {1\over 5} \left( Q_{X}-Q_{Y'} \right).
\label{ycharge}
\end{eqnarray}

There are three families of the SM fermions
whose quantum numbers under $SU(5)\times U(1)_{X}$ are
\begin{eqnarray}
F_i={\mathbf{(10, 1)}},~ {\bar f}_i={\mathbf{(\bar 5, -3)}},~
{\bar l}_i={\mathbf{(1, 5)}},
\label{smfermions}
\end{eqnarray}
where $i=1, 2, 3$. The particle assignments for the SM fermions are
\begin{eqnarray}
F_i=(Q_i, D^c_i, N^c_i)~,~~{\overline f}_i=(U^c_i, L_i)~,~~{\overline
l}_i=E^c_i~.~
\label{smparticles}
\end{eqnarray}

To break the GUT and electroweak gauge symmetries, we introduce two pairs
of Higgs fields whose quantum numbers under $SU(5)\times U(1)_X$ are
\begin{eqnarray}
H={\mathbf{(10, 1)}}~,~~{\overline{H}}={\mathbf{({\overline{10}},
-1)}}~,~~
h={\mathbf{(5, -2)}}~,~~{\overline h}={\mathbf{({\bar {5}}, 2)}}~,~\,
\label{Higgse1}
\end{eqnarray}
where $h$ and ${\overline h}$ contain the Higgs doublets
$H_d$ and $H_u$, respectively.

Moreover, the flipped $SU(5)\times U(1)_X$ models can be embedded into
$SO(10)$. Under the $SU(5)\times U(1)_X$ gauge symmetry,
the $SO(10)$ representations are decomposed as follows
\begin{eqnarray}
\mathbf{10} &=& \mathbf{(5, -2)} \oplus
\mathbf{(\overline{5}, 2)} ~,~ \\
\mathbf{{16}} &=& \mathbf{(10, 1)}
\oplus \mathbf{(\overline{5}, -3)} \oplus
\mathbf{(1, 5)} ~,~\\
\mathbf{45} &=& \mathbf{(24, 0)} \oplus \mathbf{ (1, 0)}
\oplus \mathbf{(10, -4)} \oplus \mathbf{(\overline{10}, 4)}
~.~\,
\end{eqnarray}

Third, we briefly review the
Pati-Salam model. The gauge group is
$SU(4)_C \times SU(2)_L \times SU(2)_R$ which can also be embedded
into $SO(10)$.
There are three families of the SM fermions
whose quantum numbers under $SU(4)_C \times SU(2)_L \times SU(2)_R$ are
\begin{eqnarray}
F^L_i={\mathbf{(4, 2, 1)}}~,~~ F^{Rc}_i={\mathbf{(\overline{4}, 1,
2)}}~,~\,
\end{eqnarray}
where $i=1, 2, 3$.
Also, the particle assignments for the SM fermions are
\begin{eqnarray}
F^L_i=(Q_i, L_i)~,~~F^{Rc}_i=(U^c_i, D^c_i, E^c_i, N^c_i)~.~\,
\end{eqnarray}

To break the Pati-Salam and electroweak gauge symmetries,
we introduce one pair
of Higgs fields and one bidoublet Higgs field whose
quantum numbers under $SU(4)_C \times SU(2)_L \times SU(2)_R$ are
\begin{eqnarray}
\Phi={\mathbf{(4, 1, 2)}}~,~~{\overline{\Phi}}={\mathbf{({\overline{4}},
1, 2)}}~,~~
H'={\mathbf{(1, 2, 2)}}~,~\,
\label{Higgse1}
\end{eqnarray}
where $H'$ contains one pair of the Higgs doublets
$H_d$ and $H_u$.

The Pati-Salam model can be embedded into
$SO(10)$ as well. Under $SU(4)_C\times SU(2)_L\times SU(2)_R$ gauge
symmetry,
the $SO(10)$ representations are decomposed as follows
\begin{eqnarray}
\mathbf{10} &=& \mathbf{(6, 1, 1)} \oplus
\mathbf{(1, 2, 2)} ~,~ \\
\mathbf{{16}} &=& \mathbf{(4, 2, 1)}
\oplus \mathbf{(\overline{4}, 1, 2)} ~,~\\
\mathbf{45} &=& \mathbf{(15, 1, 1)} \oplus \mathbf{ (1, 3, 1)}
\oplus \mathbf{ (1, 1, 3)}
\oplus \mathbf{(6, 2, 2)} ~.~\,
\end{eqnarray}

\section{General Gravity Mediated Supersymmetry Breaking}
\label{sec-1}
The supegravity scalar potential can be written as~\cite{mSUGRA}
\beqa {V}=M_*^4e^G\[G^i(G^{-1})^j_iG_j-3\]+\f{1}{2} {\rm
Re}\[(f^{-1})_{ab}\hat{D}^a\hat{D}^b\] ~,~\eeqa
where $M_*$ is the fundamental scale,
D-terms are
\beqa
\hat{D}^a{\equiv}-G^i(T^a)_i^j\phi_j=-\phi^{j*}(T^a)_j^iG_i~,\eeqa
and the K\"ahler function $G$ as well as its
derivatives and metric $G_i^j$ are
\beqa
G&{\equiv}&\f{K}{M_{*}^2}+\ln\(\f{W}{M_{*}^3}\)+\ln\(\f{W^*}{M_*^3}\)~,\\
G^i&=&\f{\delta G}{\de \phi_i}~,~~G_i=\f{\de G}{\de
\phi_i^*}~,~~G_i^j=\f{\de^2 G}{\de\phi^*_i\de\phi_j}~,~\,
\eeqa
where $K$ is K\"ahler potential and $W$ is superpotential.

Because the gaugino masses have been studied
previously~\cite{Li:2010xr}, we only
consider the supersymmetry breaking scalar masses and trilinear
soft terms in this paper.
To break supersymmetry, we introduce a chiral superfield $S$ in
the hidden sector whose $F$ term acquires a vacuum expectation
value (VEV), ${\it i.e}$, $\langle S \rangle = \theta^2 F_S$.
To calculate the scalar masses and trilinear
soft terms, we consider the following superpotential
and K\"ahler potential
\begin{eqnarray}
W &=& {1\over 6} y^{ijk} \phi_i \phi_j \phi_k +
\alpha {S \over {M_*}}
\left( {1\over 6} y^{ijk} \phi_i \phi_j \phi_k \right)~,~\,
\end{eqnarray}
\begin{eqnarray}
K &=& \phi_i^{\dagger} \phi_i + \beta {{S^{\dagger} S}\over {M^2_*}}
\phi_i^{\dagger} \phi_i
~,~\,
\end{eqnarray}
where $y^{ijk}$, $\alpha$, and $\beta$ are Yukawa couplings.
Thus, we obtain the universal supersymmetry breaking
scalar mass $m_0$ and
trilinear soft term $A$ of mSUGRA
\begin{eqnarray}
m^2_0~=~ \beta {{|F_S|^2}\over {M^2_*}}~,~~~ A~=~ \alpha {{F_S}\over
{M_*}}~.~\,
\end{eqnarray}

When we break the GUT gauge symmetry by giving VEV to the
Higgs field $\Phi$, we can have the general superpotential
and K\"ahler potential
\begin{eqnarray}
W &=& {1\over 6} y^{ijk} \phi_i \phi_j \phi_k
+ {1\over 6} \left( h^{ijk} {{\Phi}\over M_*} \phi_i \phi_j \phi_k
\right)
+ \alpha {S \over {M_*}}
\left( {1\over 6} y^{ijk} \phi_i \phi_j \phi_k \right) \nonumber \\
&& + \alpha' {T \over {M_*}}
\left( {1\over 6} y^{ijk} {{\Phi}\over M_*} \phi_i \phi_j \phi_k \right)
~,~\,
\end{eqnarray}
\begin{eqnarray}
K &=& \phi_i^{\dagger} \phi_i + {1\over 2} h' \phi_i^{\dagger}
\left({{\Phi}\over M_*}+
{{{\Phi}^{\dagger}} \over M_*} \right) \phi_i
+ \beta {{S^{\dagger} S}\over {M^2_*}} \phi_i^{\dagger} \phi_i
+ {1\over 2}
\beta' {{T^{\dagger} T}\over {M^2_*}} \phi_i^{\dagger}
\left({{\Phi}\over M_*}+
{{{\Phi}^{\dagger}} \over M_*} \right) \phi_i
~,~\,
\end{eqnarray}
where $h^{ijk}$, $\alpha'$, $\beta'$ and $h'$ are Yukawa couplings, and
$T$ can be $S$ or another chiral superfield with non-zero $F$ term,
${\it i.e}$, $\langle T \rangle = \theta^2 F_T$.
Therefore, after the GUT gauge symmetry is broken
by the VEV of $\Phi$, we obtain the
non-universal supersymmetry breaking scalar masses and trilinear soft
terms, which will be studied in the following.
For simplicity, we assume $h'=0$ in the following discussions
since we can redefine the fields and the SM fermion Yukawa couplings.

\section{ Scalar Masses
and Trilinear Soft Terms in the $SU(5)$ Model}
\label{sec-2}

First, we study the supersymmetry breaking scalar masses.
In order to construct gauge invariant high-dimensional operators,
we need the decompositions of
the following tensor products
\beqa {\bf \bar{5}}\otimes{\bf 5}&=&{\bf 1}\oplus {\bf
24}~,~\, \eeqa \beqa {\bf \overline{10}}\otimes {\bf 10}={\bf 1}\oplus
{\bf
24}\oplus {\bf 75}~.
\eeqa
Thus, the adjoint Higgs field can give scalar masses to both $F'_i$ and
$\overline{f}'_i$, while the Higgs field in the ${\bf 75}$ representation
can only give soft masses to $F'_i$.
The VEVs of the Higgs field $\Phi_{\bf 24}$ in the
adjoint representation are expressed as $5\times 5$ and $10\times 10$
matrices
\beqa
\langle \Phi_{\bf 24} \rangle ~=~ v {\sqrt {3\over 5}} {\rm diag}
\left(-{1\over 3}, -{1\over 3}, -{1\over 3},
{1\over 2}, {1\over 2} \right)~,~\,
\eeqa
\beqa
\langle \Phi_{\bf 24} \rangle ~=~v
\sqrt{\f{3}{5}} {\rm diag} (\underbrace{-\f{2}{3},\cdots,
-\f{2}{3}}_3,\underbrace{~\f{1}{6},\cdots,~\f{1}{6}}_{6},~1)~,
\eeqa
which are normalized to $c=1/2$ and
$c=3/2$, respectively. Thus,
we obtain the following scalar masses
\beqa m_{\tl{Q}_i}^2&=&(m_0^{U})^2+\sqrt{\f{3}{5}}\beta'_{\bf
10}\f{1}{6}(m_0^{N})^2 ~,\\
m_{\tl{U}_i^c}^2&=&(m_0^{U})^2-\sqrt{\f{3}{5}}\beta'_{\bf
10}\f{2}{3}(m_0^{N})^2 ~,\\
m_{\tl{E}_i^c}^2&=&(m_0^{U})^2+\sqrt{\f{3}{5}}\beta'_{\bf 10}(m_0^{N})^2
~,\\
m_{\tl{D}_i^c}^2&=&(m_0^{U})^2+\sqrt{\f{3}{5}}
\beta'_{\bf \bar{5}}\f{1}{3}(m_0^{N})^2 ~,\\
m_{\tl{L}_i}^2&=&(m_0^{U})^2-\sqrt{\f{3}{5}}\beta'_{\bf
\bar{5}}\f{1}{2}(m_0^{N})^2 ~,
\eeqa
where we introduced
\beqa
(m_0^{U})^2 ~\equiv~
\f{\beta}{M_*^2}F_S^*F_S~,~~(m_0^{N})^2=\f{v}{M_*^3}F_T^*F_T~.
\label{M0N-U}
\eeqa

Because the second non-universal terms are proportional to the
hypercharge for each fields,
we obtain general relations among the supersymmetry breaking scalar
masses \beqas
\f{Y_{L_i}m_{\tl{D}^c_i}^2-{Y_{D^c_i}}m_{\tl{L}_i}^2}{Y_{L_i}-Y_{D^c_i}}=\f{Y_{{U}_i^c}
m_{\tl{Q}_i}^2-{Y_{Q_i}m_{\tl{U}_i^c}^2}}{Y_{{U}_i^c}-Y_{Q_i}}
=\f{Y_{{E^c_i}}m_{\tl{Q}_i}^2-{Y_{Q_i}m_{\tl{E}_i^c}^2}}{Y_{{E}_i^c}-Y_{Q_i}}
=\f{Y_{{U}_i^c}m_{\tl{E}_i^c}^2-{Y_{E_i^c}m_{\tl{U}_i^c}^2}}{Y_{{U}_i^c}-Y_{E_i^c}}
~,\eeqas
which give the scalar mass relations at the GUT scale $M_U$
\beqa
{3m_{\tl{D}_i^c}^2+2m_{\tl{L}_i}^2}={4m_{\tl{Q}_i}^2+m_{\tl{U}_i^c}^2}
={6m_{\tl{Q}_i}^2-m_{\tl{E}_i^c}^2}={2m_{\tl{E}_i^c}^2+3m_{\tl{U}_i^c}^2}~.
\label{SMass-R}
\eeqa

Next, we consider the Higgs field $\Phi_{kl}^{[ij]}$
in the ${\bf 75}$ representation.
Because the Higgs fields $\Phi_{\bf 24}$ and $\Phi_{kl}^{[ij]}$
belong to the decomposition of the tensor product representation of
$ {\bf \overline{10}} \tm {\bf 10}$,
their VEVs must be orthogonal to each other. Thus, we obtain
the VEV of $\Phi_{kl}^{[ij]}$ in terms of the $10\times 10$ matrix
\beqa
\langle \Phi_{kl}^{[ij]} \rangle ~=~\f{v}{2\sqrt{3}}
{\rm diag}
\left(\underbrace{~1,\cdots,~1}_3,\underbrace{-1,\cdots,-1}_{6},
3\right)~.~ \,
\eeqa
So we obtain scalar masses
\beqa
m_{\tl{Q}_i}^2&=&(m_0^{U})^2-\f{\beta'_{\bf 75}}{2\sqrt{3}}(m_0^{N})^2 ~,
\nonumber \\
m_{\tl{U}_i^c}^2&=&(m_0^{U})^2+\f{\beta'_{\bf 75}}{2\sqrt{3}}(m_0^{N})^2
~, \nonumber \\
m_{\tl{E}_i^c}^2&=&(m_0^{U})^2+3\f{\beta'_{\bf 75}}{2\sqrt{3}}(m_0^{N})^2
~,\nonumber \\
m_{\tl{D}_i^c}^2&=& m_{\tl{L}_i}^2 ~=~ (m_0^{U})^2~,~
\label{SUV-75SM}
\eeqa
which respect the scalar mass relation at $M_U$
\beqa
m_{\tl{E}_i^c}^2+m_{\tl{Q}_i}^2&=&2m_{\tl{U}_i^c}^2~.
\label{SMass-R75}
\eeqa

Second, we study the supersymmetry breaking trilinear soft terms.
For simplicity, we assume that the Yukawa couplings are diagonal.
To get the possible high-dimensional operators for the trilinear soft
terms, we need to consider the decompositions of the tensor products for
the
SM fermion Yukawa coupling terms~\cite{Slansky:1981yr}
\beqa
{\bf 10}\otimes{\bf 10}\otimes{\bf
5}&=&({\bf \bar{5}} \oplus
{\bf \overline{45}}\oplus{\bf \overline{50}} )\otimes {\bf 5} \nonumber
\\
&=& ({\bf 1} \oplus {\bf 24} ) \oplus
({\bf 24}\oplus {\bf 75}\oplus{\bf 126}) \oplus
({\bf 75}\oplus {\bf 175'}) ~,~ \\
{\bf 10}\otimes{\bf \bar{5}}\otimes {\bf \bar{5}}&=&{\bf
10}\otimes({\bf \overline{10}\oplus \overline{15}})=({\bf 1 \oplus
24\oplus
75})\oplus({\bf 24 \oplus \overline{126} })~.
\eeqa
Because the Higgs fields in the ${\bf 126}$,
${\bf \overline{126}}$ and ${\bf 175'}$ do not have
the $SU(3)_C\times SU(2)_L$ singlets~\cite{Slansky:1981yr},
we do not consider them in the
following discussions.  Thus, we only consider
the Higgs fields in
the ${\bf 24}$ and ${\bf 75}$ representations.

For the Higgs field $\Phi_{\bf 24}$
in the ${\bf 24}$ representation, we consider
the following superpotential for the additional contributions
to the Yukawa coupling terms and trilinear soft terms
\beqa
W & \supset & \left(h^{Ui} \epsilon^{mnpql} (F'_i)_{mn} (F'_i)_{pq}
(h')_k
(\Phi_{\bf 24})_l^k + h^{\prime Ui}
\epsilon^{mnpkl} (F'_i)_{mn} (F'_i)_{pq} (h')_k
(\Phi_{\bf 24})_l^q \right. \nonumber \\ && \left.
+ h^{DEi} (F'_i)_{mn} (\overline{f}'_i \otimes
\overline{h}')_{Sym}^{ml} (\Phi_{\bf 24})_l^n
+ h^{\prime DEi} (F'_i)_{mn} (\overline{f}'_i \otimes
\overline{h}')_{Asym}^{ml} (\Phi_{\bf 24})_l^n \right)
\nonumber \\ &&
+ \alpha' {T \over {M_*}} \left(
y^{Ui} \epsilon^{mnpql} (F'_i)_{mn} (F'_i)_{pq} (h')_k
(\Phi_{\bf 24})_l^k
\right. \nonumber \\ && \left.
+ y^{\prime Ui}
\epsilon^{mnpkl} (F'_i)_{mn} (F'_i)_{pq} (h')_k
(\Phi_{\bf 24})_l^q
+ y^{DEi} (F'_i)_{mn} (\overline{f}'_i \otimes
\overline{h}')_{Sym}^{ml} (\Phi_{\bf 24})_l^n
\right. \nonumber \\ && \left.
+ y^{\prime DEi} (F'_i)_{mn} (\overline{f}'_i \otimes
\overline{h}')_{Asym}^{ml} (\Phi_{\bf 24})_l^n \right) ~,~\,
\eeqa
where the subscripts $Sym$ and $Asym$ denote the symmetric and
anti-symmetric products of two ${\bf
\bar{5}}$ representations.
After $\Phi_{\bf 24}$ acquires a VEV, we obtain the Yukawa coupling terms
in the superpotential
\beqa
W & \supset & {v\over {M_*}} \sqrt{\f{3}{5}} \left(
-2 h^{Ui} Q_i U_i^c H_u - h^{\prime Ui} Q_i U_i^c H_u
\right. \nonumber \\ && \left.
-{1\over 6} h^{\prime DEi} Q_i D^c_i H_d
- h^{\prime DEi} L_i E^c_i H_d
\right)~.~ \,
\eeqa
We also obtain the supersymmetry breaking trilinear soft terms
\beqa
-{\cal L} & \supset & \alpha' {{F_T v} \over {M^2_*}} \sqrt{\f{3}{5}}
\left(-2 y^{Ui} \tl{Q}_i \tl{U}_i^c H_u
- y^{\prime Ui} \tl{Q}_i \tl{U}_i^c H_u
\right. \nonumber \\ && \left.
-{1\over 6} y^{\prime DEi} \tl{Q}_i \tl{D}^c_i H_d
- y^{\prime DEi} \tl{L}_i \tl{E}^c_i H_d
\right)~.~ \,
\eeqa
As a double check, we can obtain these results by choosing the VEVs of
${\bf 24}$ dimensional Higgs field
$\Phi_{\bf 24}$ as appropriate $5\tm 5$ and
$10\tm 10 $ matrices.

We can write the VEV of the
${\bf 75}$ dimensional Higgs field $\Phi_{jl}^{[ik]}$
as~\cite{Ellis:1985jn}
\beqa
\langle \Phi^{[ik]}_{jl} \rangle ~=~\f{v}{2\sqrt{3}}
\[\Delta_{cj}^{[i}\Delta_{cl}^{k]}
+2\Delta_{wj}^{[i}\Delta_{wl}^{k]}
-\f{1}{2}\delta^{[i}_{j}\delta^{k]}_{l}\]~,
\eeqa
where
\beqa\Delta_c~=~{\rm diag}(~1,~1,~1,~0,~0)~,~~
\Delta_w~=~{\rm diag}(~0,~0,~0,~1,~1)~.
\eeqa
We consider
the following superpotential for the additional contributions
to the Yukawa coupling terms and trilinear soft terms
\beqa
W & \supset & \left(h^{Ui} \epsilon^{mnpjl} (F'_i)_{mn} (F'_i)_{pq}
(h')_k
\Phi^{[qk]}_{jl} + h^{\prime Ui}
\epsilon^{jlpqk} (F'_i)_{mn} (F'_i)_{pq} (h')_k
\Phi^{[mn]}_{jl}
\right. \nonumber \\ && \left.
+ h^{DEi} (F'_i)_{mn} (\overline{f}'_i)^p (\overline{h}')^q
\Phi^{[mn]}_{pq} \right)
+ \alpha' {T \over {M_*}} \left(
y^{Ui} \epsilon^{mnpjl} (F'_i)_{mn} (F'_i)_{pq} (h')_k
\Phi^{[qk]}_{jl}
\right. \nonumber \\ && \left.
+ y^{\prime Ui}
\epsilon^{jlpqk} (F'_i)_{mn} (F'_i)_{pq} (h')_k
\Phi^{[mn]}_{jl}
+ y^{DEi} (F'_i)_{mn} (\overline{f}'_i)^p (\overline{h}')^q
\Phi^{[mn]}_{pq} \right)
~.~\,
\eeqa

After $\Phi^{[ik]}_{jl}$ acquires a VEV, we obtain the Yukawa coupling
terms
in the superpotential
\beqa
W & \supset & {v\over {M_*}} \f{1}{2 {\sqrt {3}}} \left(
- h^{\prime DEi} Q_i D^c_i H_d
+3 h^{\prime DEi} L_i E^c_i H_d
\right)~,~ \,
\eeqa
and the supersymmetry breaking trilinear soft terms 
\beqa
-{\cal L} & \supset & \alpha' {{F_T v} \over {M^2_*}} \f{1}{2 {\sqrt
{3}}}
\left(
- y^{\prime DEi} \tl{Q}_i \tl{D}^c_i H_d
+3 y^{\prime DEi} \tl{L}_i \tl{E}^c_i H_d
\right)~.~ \,
\eeqa
These results can also be obtained by considering the
VEV of ${\bf 75}$ dimensional Higgs field as an
appropriate $10\tm 10 $ matrix.
Due to the arbitrariness of the coefficients in the Yukawa
coupling terms and the trilinear soft terms, we will not discuss the
relations
among the trilinear soft terms.

\section{ Scalar Masses in the $SO(10)$ Model}
\label{sec-3}

In order to calculate the scalar masses,
we need to decompose the tensor product
of ${\bf \overline{16}}\otimes{\bf 16}$ which gives
\beqa {\bf \overline{16}}\otimes{\bf 16}={\bf
1}\oplus{\bf 45}\oplus{\bf 210} ~.~\,
\eeqa
Thus we need to consider the Higgs fields in the
${\bf 45}$ and ${\bf 210}$ representations to determine the scalar
masses.
The $SO(10)$ gauge symmetry can be broken down to the Pati-Salam
$SU(4)_C \times SU(2)_L \times SU(2)_R$ gauge symmetry by the
Higgs fields in the ${\bf 45}$, ${\bf 210}$, and ${\bf 770}$
representations, and can be (further) broken down to the
$SU(3)_C\times SU(2)_L \times SU(2)_R \times U(1)_{B-L}$ gauge
symmetry by the Higgs field in the
$({\bf 15}, {\bf 1}, {\bf 1})$ component
of the $SU(4)_C\times SU(2)_L \times SU(2)_R$
under the ${\bf 45}$ and ${\bf 210}$ representations.
In addition,
the $SO(10)$ gauge symmetry can be broken down to the Georgi-Glashow
$SU(5)\times U(1)'$ and flipped $SU(5)\times U(1)_X$ gauge
symmetries by the
Higgs fields in the ${\bf 45}$ and ${\bf 210}$
representations, and can be (further) broken down to the
$SU(3)_C\times SU(2)_L \times U(1)_1 \times U(1)_2$ gauge
symmetries by the Higgs field in the $({\bf 24}, {\bf 0})$ component
of the $SU(5)\times U(1)$ under the ${\bf 45}$ representation, or by
the $({\bf 24}, {\bf 0})$ or $({\bf 75}, {\bf 0})$ component
under the ${\bf 210}$ representation.
Thus, in the following,
we consider these breaking chains.

\subsection{The Pati-Salam Model}

From the decomposition of the ${\bf
16}$ dimensional spinor representation under
the $SU(4)_C \times SU(2)_L \times SU(2)_R$ gauge symmetry,
we obtain the VEV
(of the $({\bf 1,1,1})$ component) of the
$210$ dimensional Higgs field $\Phi_{\bf 210}$ in
terms of the $16\tm 16$ matrix
\beqa
<\Phi_{\bf 210}>=\f{v}{2\sqrt{2}}
{\rm diag}
(\underbrace{~1,\cdots,~1}_8,\underbrace{-1,\cdots,-1}_8) ~,~\,
\eeqa
with the normalization $c=2$. From this we get the scalar masses
\beqa
M^2(\tilde{F}^L_i)&=& (m_0^U)^2 + \f{v}{2\sqrt{2}} \beta'_{\bf 210}
\f{v |F_T|^2}{M_*^3} ~,\\
M^2(\tilde{F}_i^{Rc}) &=& (m_0^U)^2 -\f{v}{2\sqrt{2}} \beta'_{\bf
210}\f{v
|F_T|^2}{M_*^3}~. \eeqa
In components, we have
\beqa
m_{\tl{Q}_i}^2&=&(m_0^{U})^2 + \f{v}{2\sqrt{2}}\beta'_{\bf 210}
\f{v |F_T|^2}{M_*^3}~, \nonumber \\
m_{\tl{U}_i^c}^2&=&(m_0^{U})^2 - \f{v}{2\sqrt{2}}\beta'_{\bf 210}\f{v
|F_T|^2}{M_*^3} ~, \nonumber \\
m_{\tl{E}_i^c}^2&=&(m_0^{U})^2 - \f{v}{2\sqrt{2}}\beta'_{\bf 210}\f{v
|F_T|^2}{M_*^3} ~, \nonumber \\
m_{\tl{D}_i^c}^2&=&(m_0^{U})^2 - \f{v}{2\sqrt{2}} \beta'_{\bf 210}\f{v
|F_T|^2}{M_*^3}~, \nonumber \\
m_{\tl{L}_i}^2&=&(m_0^{U})^2 + \f{v}{2\sqrt{2}} \beta'_{\bf 210}\f{v
|F_T|^2}{M_*^3}~.
\label{PSsm-210}
\eeqa

\subsection{The $SU(3)_C\tm SU(2)_L\tm SU(2)_R\tm U(1)_{B-L}$ Model}

The $SO(10)$ gauge symmetry can be broken down to
the $SU(3)_C\tm SU(2)_L\tm SU(2)_R\tm U(1)_{B-L}$ symmetry
by giving VEVs to the $({\bf 15,1,1})$ components
of the Higgs field
in the ${\bf 45}$ and ${\bf 210}$ representations
of $SU(4)_C\times SU(2)_L \times SU(2)_R$.
The decomposition of ${\bf 16}$ under the $SU(3)_C\tm SU(2)_L\tm
SU(2)_R\tm U(1)_{B-L}$ group is
\beqa {\bf
16}={\bf (3,2,1, {1/6})}\oplus {\bf (1,2,1, {-1/2})} \oplus
{\bf (\bar{3},1,\bar{2}, {-1/6})}\oplus {\bf
(1,1,\bar{2}, {1/2})}~. \eeqa

First, let us consider the Higgs field $\Phi_{\bf 45}$ in the
${\bf 45}$ representation. The VEV of $\Phi_{\bf 45}$
can be written in terms of a $16\tm 16$ matrix as follows
\beqa
\langle \Phi_{\bf 45} \rangle ~=~\f{v}{2\sqrt{6}}
{\rm diag}(\underbrace{~1,~1,~1,-3}_2,\underbrace{-1,-1,-1,~3}_2)~, \eeqa
which is normalized as $c=2$. Thus, the scalar masses are
\beqa
m_{\tl{Q}_i}^2&=&(m_0^{U})^2+\f{v}{2\sqrt{6}} \beta'_{\bf 45}\f{v
|F_T|^2}{M_*^3}~,
\nonumber \\
m_{\tl{U}_i^c}^2&=&(m_0^{U})^2-\f{v}{2\sqrt{6}} \beta'_{\bf 45}\f{v
|F_T|^2}{M_*^3} ~,\nonumber \\
m_{\tl{E}_i^c}^2&=&(m_0^{U})^2+\f{3v}{2\sqrt{6}} \beta'_{\bf 45}\f{v
|F_T|^2}{M_*^3} ~,\nonumber \\
m_{\tl{D}_i^c}^2&=&(m_0^{U})^2-\f{v}{2\sqrt{6}} \beta'_{\bf 45}\f{v
|F_T| ^2}{M_*^3}~,\nonumber \\
m_{\tl{L}_i}^2&=&(m_0^{U})^2-\f{3v}{2\sqrt{6}} \beta'_{\bf 45}\f{v
|F_T| ^2}{M_*^3}~.
\eeqa

Second, we consider the Higgs field $\Phi_{\bf 210}$ in the
${\bf 210}$ representation. The VEV of $\Phi_{\bf 210}$
in terms of a $16\tm 16$ matrix is
\beqa
\langle \Phi_{\bf 210} \rangle ~=~
\f{v}{2\sqrt{6}} {\rm diag}(\underbrace{~1,~1,~1,-3}_4)~, \eeqa
which is normalized as $c=2$. Thus, the scalar masses are
\beqa
m_{\tl{Q}_i}^2&=&(m_0^{U})^2+\f{v}{2\sqrt{6}} \beta'_{\bf 210}\f{v
|F_T|^2}{M_*^3}~,
\nonumber \\
m_{\tl{U}_i^c}^2&=&(m_0^{U})^2+\f{v}{2\sqrt{6}} \beta'_{\bf 210}\f{v
|F_T|^2}{M_*^3} ~, \nonumber \\
m_{\tl{E}_i^c}^2&=&(m_0^{U})^2-\f{3v}{2\sqrt{6}} \beta'_{\bf 210}\f{v
|F_T|^2}{M_*^3} ~, \nonumber \\
m_{\tl{D}_i^c}^2&=&(m_0^{U})^2+\f{v}{2\sqrt{6}} \beta'_{\bf 210}\f{v
|F_T|^2}{M_*^3}~,\nonumber \\
m_{\tl{L}_i}^2&=&(m_0^{U})^2-\f{3v}{2\sqrt{6}} \beta'_{\bf 210}\f{v
|F_T|^2}{M_*^3}~.
\label{SOT-210}
\eeqa

\subsection{ The Georgi-Glashow $SU(5)\tm U(1)'$ and Flipped $SU(5)\times
U(1)_X$ Models}

The $SO(10)$ gauge symmetry can also be broken down to
the $SU(5)\tm U(1)$ gauge symmetry by the ${\bf 45}$ and
${\bf 210}$ dimensional Higgs fields $\Phi_{\bf 45}$ and
$\Phi_{\bf 210}$. The decomposition of the ${\bf 16}$ spinor
representation under $SU(5)\tm U(1)$ is
\beqa {\bf
16}={\bf (10, ~1)\oplus (\bar{5},~-3)\oplus (1,~5)} ~.
\label{SOT-GGFSUV}
\eeqa

First, we consider the Higgs field $\Phi_{\bf 45}$.
From Eq.~(\ref{SOT-GGFSUV}), we obtain the VEV of $\Phi_{\bf 45}$
in terms of a $16\tm 16$ matrix
\beqa
\langle \Phi_{\bf 45} \rangle ~=~ \f{v}{2\sqrt{10}}
{\rm diag}(\underbrace{-3,\cdots,-3}_5,\underbrace{1,\cdots,1}_{10},5)~,
\eeqa
which is normalized as $c=2$. Consequently, we obtain the scalar masses
in
the Georgi-Glashow $SU(5)\tm U(1)'$
and flipped $SU(5)\times U(1)_X$ models:
\begin{itemize}
\item The Georgi-Glashow $SU(5)\tm U(1)'$ Model
\beqa
M^2(\tl{F}'_i)&=&(m_0^U)^2+\beta'_{\bf 45}\f{v |F_T|^2}{2\sqrt{10}M_*^3}
~, \nonumber \\
M^2(\tl{\overline{f}}'_i)&=&(m_0^U)^2-3 \beta'_{\bf 45}\f{v
|F_T|^2}{2\sqrt{10}M_*^3}~, \nonumber \\
M^2(\tl{N}^c_i)&=&(m_0^U)^2 + 5 \beta'_{\bf 45}\f{v
|F_T|^2}{2\sqrt{10}M_*^3}~.
\eeqa
In components, we have
\beqa
m_{\tl{Q}_i}^2&=&(m_0^{U})^2 + \beta'_{\bf 45}\f{v
|F_T|^2}{2\sqrt{10}M_*^3}~, \nonumber \\
m_{\tl{U}_i^c}^2&=&(m_0^{U})^2 + \beta'_{\bf 45}\f{v
|F_T|^2}{2\sqrt{10}M_*^3}~, \nonumber \\
m_{\tl{E}_i^c}^2&=&(m_0^{U})^2 + \beta'_{\bf 45}\f{v
|F_T|^2}{2\sqrt{10}M_*^3}, \nonumber \\
m_{\tl{D}_i^c}^2&=&(m_0^{U})^2 -3 \beta'_{\bf 45}\f{v
|F_T|^2}{2\sqrt{10}M_*^3}~, \nonumber \\
m_{\tl{L}_i}^2&=&(m_0^{U})^2 - 3 \beta'_{\bf 45}\f{v
|F_T|^2}{2\sqrt{10}M_*^3}~.\eeqa
In this paper, we will not consider the scalar masses
for right-handed sneutrinos because the heavy Majorana neutrino
masses will give the dominant contributions.

\item The Flipped $SU(5)\times U(1)_X$ Model
\beqa
M^2(\tl{F}_i)&=&(m_0^U)^2+\beta'_{\bf 45}\f{v |F_T|^2}{2\sqrt{10}M_*^3}
~, \nonumber \\
M^2(\tl{\overline{f}}_i)&=&(m_0^U)^2-3 \beta'_{\bf 45}\f{v
|F_T|^2}{2\sqrt{10}M_*^3}~, \nonumber \\
M^2(\tl{\overline{l}}_i)&=&(m_0^U)^2 + 5 \beta'_{\bf 45}\f{v
|F_T|^2}{2\sqrt{10}M_*^3}~.
\eeqa
In components, this gives
\beqa
m_{\tl{Q}_i}^2&=&(m_0^{U})^2 +\beta'_ {\bf 45}\f{v
|F_T|^2}{2\sqrt{10}M_*^3}~, \nonumber \\
m_{\tl{U}_i^c}^2&=&(m_0^{U})^2 - 3 \beta'_{\bf 45}\f{v
|F_T|^2}{2\sqrt{10}M_*^3}~, \nonumber \\
m_{\tl{E}^c_i}^2&=&(m_0^{U})^2 + 5 \beta'_{\bf 45}\f{v
|F_T|^2}{2\sqrt{10}M_*^3}~, \nonumber \\
m_{\tl{D}_i^c}^2&=&(m_0^{U})^2 + \beta'_{\bf 45}\f{v
|F_T|^2}{2\sqrt{10}M_*^3}~, \nonumber \\
m_{\tl{L}_i}^2&=&(m_0^{U})^2- 3 \beta'_{\bf 45}\f{v
|F_T|^2}{2\sqrt{10}M_*^3}~.\eeqa
\end{itemize}





Second, let us consider the Higgs field $\Phi_{\bf 210}$.
Because the VEVs of $\Phi_{\bf 45}$ and $\Phi_{\bf 210}$
are orthogonal to each other, we obtain the VEV of $\Phi_{\bf 210}$
in terms of the $16\tm 16$ matrix
\beqa <\Phi>=\f{v}{2\sqrt{5}}
{\rm
diag}(\underbrace{~1,\cdots,~1}_5,\underbrace{-1,\cdots,-1}_{10},~5)~,
\eeqa
which is normalized as $c=2$. From this, we obtain the
scalar masses in
the Georgi-Glashow $SU(5)\tm U(1)'$
and flipped $SU(5)\times U(1)_X$ models:
\begin{itemize}
\item The Georgi-Glashow $SU(5)\tm U(1)'$ Model
\beqa
M^2(\tl{F}'_i)&=&(m_0^U)^2-\beta'_{\bf 210}\f{v |F_T|^2}{2\sqrt{5}M_*^3}
~, \nonumber \\
M^2(\tl{\overline{f}}'_i)&=&(m_0^U)^2 + \beta'_{\bf 210}\f{v
|F_T|^2}{2\sqrt{5}M_*^3}~, \nonumber \\
M^2(\tl{N}^c_i)&=&(m_0^U)^2 + 5 \beta'_{\bf 210}\f{v
|F_T|^2}{2\sqrt{5}M_*^3}~.
\eeqa
In components
\beqa
m_{\tl{Q}_i}^2&=&(m_0^{U})^2 - \beta'_{\bf 210}\f{v
|F_T|^2}{2\sqrt{5}M_*^3}~, \nonumber \\
m_{\tl{U}_i^c}^2&=&(m_0^{U})^2 - \beta'_{\bf 210}\f{v
|F_T|^2}{2\sqrt{5}M_*^3}~, \nonumber \\
m_{\tl{E}_i^c}^2&=&(m_0^{U})^2 - \beta'_{\bf 210}\f{v
|F_T|^2}{2\sqrt{5}M_*^3}, \nonumber \\
m_{\tl{D}_i^c}^2&=&(m_0^{U})^2 + \beta'_{\bf 210}\f{v
|F_T|^2}{2\sqrt{5}M_*^3}~, \nonumber \\
m_{\tl{L}_i}^2&=&(m_0^{U})^2 + \beta'_{\bf 210}\f{v
|F_T|^2}{2\sqrt{5}M_*^3}~.\eeqa

\item The Flipped $SU(5)\times U(1)_X$ Model
\beqa
M^2(\tl{F}_i)&=&(m_0^U)^2 - \beta'_{\bf 210}\f{v |F_T|^2}{2\sqrt{5}M_*^3}
~, \nonumber \\
M^2(\tl{\overline{f}}_i)&=&(m_0^U)^2 + \beta'_{\bf 210}\f{v
|F_T|^2}{2\sqrt{5}M_*^3}~, \nonumber \\
M^2(\tl{\overline{l}}_i)&=&(m_0^U)^2 + 5 \beta'_{\bf 210}\f{v
|F_T|^2}{2\sqrt{5}M_*^3}~.
\eeqa
In components
\beqa
m_{\tl{Q}_i}^2&=&(m_0^{U})^2 - \beta'_ {\bf 210}\f{v
|F_T|^2}{2\sqrt{5}M_*^3}~, \nonumber \\
m_{\tl{U}_i^c}^2&=&(m_0^{U})^2 + \beta'_{\bf 210}\f{v
|F_T|^2}{2\sqrt{5}M_*^3}~, \nonumber \\
m_{\tl{E}^c_i}^2&=&(m_0^{U})^2 + 5 \beta'_{\bf 210}\f{v
|F_T|^2}{2\sqrt{5}M_*^3}~, \nonumber \\
m_{\tl{D}_i^c}^2&=&(m_0^{U})^2 - \beta'_{\bf 210}\f{v
|F_T|^2}{2\sqrt{5}M_*^3}~, \nonumber \\
m_{\tl{L}_i}^2&=&(m_0^{U})^2 + \beta'_{\bf 210}\f{v
|F_T|^2}{2\sqrt{5}M_*^3}~.
\label{FSUV-210}
\eeqa
\end{itemize}



\subsection{ The $SU(3)_C\tm SU(2)_L\tm U(1) _1\tm U(1)_2$ Model}

The $SO(10)$ gauge symmetry can also be broken down
to $SU(3)_C\tm SU(2)_L\tm U(1)_1 \tm U(1)_{2}$
by VEVs of the ${\bf (24, 0)}$ component Higgs field
which is in the ${\bf 45}$ representation under $SU(5)\times U(1)$,
or by the ${\bf (24, 0)}$ or ${\bf (75, 0)}$ component Higgs fields
in the ${\bf 210}$ representation.

First, we consider the Higgs field $\Phi_{\bf 45}$ in the
${\bf 45}$ representation. The VEV of $\Phi_{\bf 45}$ is
\beqa
\langle \Phi_{\bf 45} \rangle
~=~v \sqrt{\f{3}{5}} {\rm diag}(~\f{1}{3},~\f{1}{3},~\f{1}{3},-\f{1}{2},
-\f{1}{2},\underbrace{~\f{1}{6},\cdots,~\f{1}{6}}_6,-\f{2}{3},
-\f{2}{3},-\f{2}{3},~1,~0)~,
\eeqa
which is normalized to $c=2$.

Thus, we obtain the scalar masses in
the Georgi-Glashow $SU(5)\tm U(1)'$
and flipped $SU(5)\times U(1)_X$ models:
\begin{itemize}
\item The Georgi-Glashow $SU(5)\tm U(1)'$ Model
\beqa m_{\tl{Q}_i}^2&=&(m_0^{U})^2+\sqrt{\f{3}{5}} \beta'_{\bf 45}
\f{1}{6}(m_0^{N})^2 ~, \nonumber \\
m_{\tl{U}_i^c}^2&=&(m_0^{U})^2-\sqrt{\f{3}{5}} \beta'_{\bf 45}
\f{2}{3}(m_0^{N})^2 ~, \nonumber \\
m_{\tl{E}_i^c}^2&=&(m_0^{U})^2+\sqrt{\f{3}{5}} \beta'_{\bf 45}
(m_0^{N})^2~, \nonumber \\
m_{\tl{D}^c_i}^2&=&(m_0^{U})^2+\sqrt{\f{3}{5}} \beta'_{\bf 45}
\f{1}{3}(m_0^{N})^2 ~, \nonumber \\
m_{\tl{L}_i}^2&=&(m_0^{U})^2-\sqrt{\f{3}{5}} \beta'_{\bf 45}
\f{1}{2}(m_0^{N})^2 ~,~\,
\eeqa
where $(m_0^{U})^2$ and $(m_0^{N})^2$ are given in Eq.~(\ref{M0N-U}).

\item The Flipped $SU(5)\times U(1)_X$ Model
\beqa m_{\tl{Q}_i}^2&=&(m_0^{U})^2+\sqrt{\f{3}{5}} \beta'_{\bf 45}
\f{1}{6}(m_0^{N})^2 ~, \nonumber \\
m_{\tl{U}_i^c}^2&=&(m_0^{U})^2+\sqrt{\f{3}{5}} \beta'_{\bf 45}
\f{1}{3}(m_0^{N})^2 ~,\\
m_{\tl{E}_i^c}^2&=&(m_0^{U})^2~, \nonumber \\
m_{\tl{D}_i^c}^2&=&(m_0^{U})^2-\sqrt{\f{3}{5}} \beta'_{\bf 45}
\f{2}{3}(m_0^{N})^2 ~, \nonumber \\
m_{\tl{L}_i}^2&=&(m_0^{U})^2-\sqrt{\f{3}{5}} \beta'_{\bf 45}
\f{1}{2}(m_0^{N})^2 ~.
\eeqa
\end{itemize}

Second, we consider the Higgs field $\Phi^{\bf 24}_{\bf 210}$ in the
${\bf (24, 0)}$ component of the ${\bf 210}$ representation
that acquires a VEV as follows
\beqa
\langle \Phi^{\bf 24}_{\bf 210} \rangle
~=~\f{v}{\sqrt{5}}{\rm diag}(-1,-1,-1,~\f{3}{2},~\f{3}{2},
\underbrace{~\f{1}{6},\cdots,~\f{1}{6}}_6,-\f{2}{3},-\f{2}{3},-\f{2}{3},~1,~0)~,
\eeqa
which is normalized to $c=2$.
From this, we obtain the scalar masses in
the Georgi-Glashow $SU(5)\tm U(1)'$
and the flipped $SU(5)\times U(1)_X$ models:

\begin{itemize}
\item The Georgi-Glashow $SU(5)\tm U(1)'$ Model

\beqa
m_{\tl{Q}_i}^2&=&(m_0^{U})^2+\f{1}{\sqrt{5}} \beta^{\prime {\bf 24}}_{\bf
210}
\f{1}{6}(m_0^{N})^2 ~, \nonumber \\
m_{\tl{U}^c_i}^2&=&(m_0^{U})^2-\f{1}{\sqrt{5}} \beta^{\prime {\bf
24}}_{\bf 210}
\f{2}{3}(m_0^{N})^2 ~, \nonumber \\
m_{\tl{E}^c_i}^2&=&(m_0^{U})^2+\f{1}{\sqrt{5}} \beta^{\prime {\bf
24}}_{\bf 210}
(m_0^{N})^2~, \nonumber \\
m_{\tl{D}^c_i}^2&=&(m_0^{U})^2-\f{1}{\sqrt{5}} \beta^{\prime {\bf
24}}_{\bf 210}
(m_0^{N})^2 ~, \nonumber \\
m_{\tl{L}_i}^2&=&(m_0^{U})^2+\f{1}{\sqrt{5}} \beta^{\prime {\bf 24}}_{\bf
210}
\f{3}{2} (m_0^{N})^2 ~.
\eeqa
\item The Flipped $SU(5)\times U(1)_X$ Model

\beqa
m_{\tl{Q}_i}^2&=&(m_0^{U})^2+\f{1}{\sqrt{5}} \beta^{\prime {\bf 24}}_{\bf
210}
\f{1}{6}(m_0^{N})^2 ~, \nonumber \\
m_{\tl{U}^c_i}^2&=&(m_0^{U})^2-\f{1}{\sqrt{5}} \beta^{\prime {\bf
24}}_{\bf 210}
(m_0^{N})^2 ~, \nonumber \\
m_{\tl{E}^c_i}^2&=&(m_0^{U})^2~, \nonumber \\
m_{\tl{D}^c_i}^2&=&(m_0^{U})^2-\f{1}{\sqrt{5}} \beta^{\prime {\bf
24}}_{\bf 210}
\f{2}{3}(m_0^{N})^2 ~, \nonumber \\
m_{\tl{L}_i}^2&=&(m_0^{U})^2+\f{1}{\sqrt{5}} \beta^{\prime {\bf 24}}_{\bf
210}
\f{3}{2}(m_0^{N})^2 ~.
\eeqa
\end{itemize}

Third, we consider that the ${\bf (75, 0)}$ component
Higgs field $\Phi^{\bf 75}_{\bf 210}$ in the ${\bf 210}$ representation
acquires a VEV as follows
\beqa
\langle \Phi^{\bf 75}_{\bf 210} \rangle
~=~\f{v}{3}{\rm
diag}(~0,~0,~0,~0,~0,\underbrace{-1,\cdots,-1}_6,~1,~1,~1,~3,~0)~,~\,
\eeqa
which is normalized to $c=2$.
From this, we obtain the following scalar masses in
the Georgi-Glashow $SU(5)\tm U(1)'$
and the flipped $SU(5)\times U(1)_X$ models:

\begin{itemize}
\item The Georgi-Glashow $SU(5)\tm U(1)'$ Model

\beqa m_{\tl{Q}_i}^2&=&(m_0^{U})^2-\f{1}{3} \beta^{\prime {\bf 75}}_{\bf
210}
(m_0^{N})^2 ~, \nonumber \\
m_{\tl{U}^c_i}^2&=&(m_0^{U})^2+\f{1}{3} \beta^{\prime {\bf 75}}_{\bf 210}
(m_0^{N})^2 ~, \nonumber \\
m_{\tl{E}^c_i}^2&=&(m_0^{U})^2+ \beta^{\prime {\bf 75}}_{\bf 210}
(m_0^{N})^2~, \nonumber \\
m_{\tl{D}^c_i}^2&=&(m_0^{U})^2~, \nonumber \\
m_{\tl{L}_i}^2&=&(m_0^{U})^2~.
\eeqa

\item The Flipped $SU(5)\times U(1)_X$ Model

\beqa m_{\tl{Q}_i}^2&=&(m_0^{U})^2-\f{1}{3} \beta^{\prime {\bf 75}}_{\bf
210}
(m_0^{N})^2 ~, \nonumber \\
m_{\tl{U}^c_i}^2&=&(m_0^{U})^2~, \nonumber \\
m_{\tl{E}^c_i}^2&=&(m_0^{U})^2~, \nonumber \\
m_{\tl{D}^c_i}^2&=&(m_0^{U})^2+\f{1}{3} \beta^{\prime {\bf 75}}_{\bf 210}
(m_0^{N})^2 ~, \nonumber \\
m_{\tl{L}_i}^2&=&(m_0^{U})^2~.
\eeqa
\end{itemize}

\section{ The Yukawa Coupling Terms and
Trilinear Soft Terms in the $SO(10)$ Model}

There are several kinds of the
renormalizable Yukawa coupling terms for the SM fermions
in the $SO(10)$ model.
For example, we can use ${\bf 120}$ or ${\bf 126}$ Higgs fields
to obtain reasonable SM fermion masses and mixings.
In this paper we choose the simplest Higgs field $H_{\bf 10}$ in the
$SO(10)$
fundamental representation.
To obtain the non-renormalizable contributions to the Yukawa coupling
terms and
trilinear soft terms, we need to know the decompositions of the tensor
product
${\bf 16}\otimes{\bf 16}\otimes {\bf 10}$~\cite{Slansky:1981yr}
\beqa
{\bf 16}\otimes{\bf 16}&=&{\bf 10\oplus 120\oplus 126}~,\\
{\bf 16}\otimes{\bf 16}\otimes {\bf 10}&=&{\bf ( 1 \oplus 45\oplus
54)\oplus ( 45\oplus 210\oplus 945 )\oplus(210 \oplus 1050 ) }~.~\,
\eeqa
Because the ${\bf 945}$ and ${\bf 1050}$ representations do not
have $SU(5)\times U(1)$ or $SU(4)_C\times SU(2)_L \times SU(2)_R$
singlets~\cite{Slansky:1981yr},
we only consider the Higgs fields in the ${\bf 45}$,
${\bf 54}$ and ${\bf 210}$ representations.



\subsection{The Pati-Salam Model}

The $SO(10)$ gauge symmetry can be broken down
to the Pati-Salam $SU(4)_C\times SU(2)_L \times SU(2)_R$
gauge symmetry
by giving VEVs to the Higgs fields in the
${\bf 54}$ and ${\bf 210}$ representations.

For the Higgs field $\Phi_{\bf 54}$ in the ${\bf 54}$ representation,
we can write the VEV in terms of a
${\bf 10\tm 10}$ matrix
\beqa
\langle \Phi_{\bf 54} \rangle =\f{v}{2\sqrt{15}}
{\rm diag}(\underbrace{~2,\cdots,~2}_6,\underbrace{-3,\cdots,-3}_4) ~,~\,
\eeqa
which is normalized to $c=1$.

To calculate the additional contributions to the Yukawa coupling terms
and trilinear soft terms, we consider the following
superpotential
\beqa
W & \supset& {1\over {M_*}}
h^i ({\bf 16_i\otimes 16_i})_{\bf
10}^m (\Phi_{\bf 54})_{mn}{\bf 10}^n
+ \alpha' {T\over {M^2_*}}
y^i ({\bf 16_i\otimes 16_i})_{\bf
10}^m (\Phi_{\bf 54})_{mn}{\bf 10}^n ~.~\,
\eeqa
After $\Phi_{\bf 54}$ acquires a VEV, we obtain the additional
contributions to the Yukawa coupling terms
\beqa
W & \supset& - h^i \f{3
v}{\sqrt{15}M_*}\left[{Q}_i{U}_i^cH_u+{L}_i{N}^c_i H_u
+ {Q}_iD_i^c H_d + {L}_i{E_i^c} H_d \right]~.~\,
\eeqa
The extra supersymmetry breaking trilinear soft terms are
\beqa
-{\cal L} & \supset & - y^i \f{3F_T v}{\sqrt{15}M_*^2}
\left[\tl{Q}_i \tl{U}_i^c H_u+\tl{L}_i \tl{N}^c_i H_u
+ \tl{Q}_i \tl{D}_i^c H_d + \tl{L}_i \tl{E}_i^c H_d \right]~.~\,
\eeqa

For the Higgs field $\Phi_{\bf 210}$ in the ${\bf 210}$ representation,
we can write the VEV in terms of a
${\bf 16 \tm 16}$ matrix
\beqa
\langle \Phi_{\bf 210} \rangle
=\f{v}{2\sqrt{2}}
{\rm diag}(\underbrace{~1,\cdots,~1}_8,\underbrace{-1,\cdots,-1}_8) ~,~\,
\eeqa
which is normalized to $c=2$.
We consider the following superpotential
\beqa
W & \supset& {1\over {M_*}}
\left[ h^i ({\bf 16_i\otimes 16_i})_{\bf
120}^{mnl} (\Phi_{\bf 210})_{mnlk}{\bf 10}^k
+ h^{\prime i} ({\bf 16_i\otimes 16_i})_{\bf
126}^{mnlpq} (\Phi_{\bf 210})_{mnlp}{\bf 10}_q
\right]
\nonumber \\ &&
+ \alpha' {T\over {M^2_*}}
\left[ y^i ({\bf 16_i\otimes 16_i})_{\bf
120}^{mnl} (\Phi_{\bf 210})_{mnlk}{\bf 10}^k
\right. \nonumber \\ && \left.
+ y^{\prime i} ({\bf 16_i\otimes 16_i})_{\bf
126}^{mnlpq} (\Phi_{\bf 210})_{mnlp}{\bf 10}_q
\right] ~.~\,
\eeqa
We can show that the above superpotential will not contribute to
the SM fermion Yukawa coupling terms and trilinear soft terms.

\subsection{The $SU(3)_C\tm SU(2)_L\tm SU(2)_R\tm U(1)_{B-L}$ Model}

The $SO(10)$ gauge symmetry can also be broken down to
$SU(3)_C\tm SU(2)_L\tm SU(2)_R\tm U(1)_{B-L}$ by giving
VEVs to the $({\bf 15,~1,~1})$ components of the Higgs fields
in the ${\bf 45}$ and ${\bf 210}$ representations under
$SU(4)_C\times SU(2)_L \times SU(2)_R$.

For the Higgs field $\Phi_{\bf 45}$ in the ${\bf 45}$ representation,
we can write the VEV
in terms of a $10\tm 10$ matrix as follows
\beqa
\langle \Phi_{\bf 45} \rangle =\f{v}{2\sqrt{6}}
{\rm diag}
(\underbrace{~2,\cdots,~2}_3,\underbrace{-2,\cdots,-2}_3,\underbrace{~0,\cdots,~0}_4)~,
\eeqa
which is normalized as $c=1$.

To calculate the additional contributions to the Yukawa coupling terms
and trilinear soft terms, we consider the following
superpotential
\beqa
W & \supset& {1\over {M_*}}
\left[ h^i ({\bf 16_i\otimes 16_i})_{\bf
10}^{m} (\Phi_{\bf 45})_{mn}{\bf 10}^n
+ h^{\prime i} ({\bf 16_i\otimes 16_i})_{\bf
120}^{mnl} (\Phi_{\bf 45})_{mn}{\bf 10}_l
\right]
\nonumber \\ &&
+ \alpha' {T\over {M^2_*}}
\left[ y^i ({\bf 16_i\otimes 16_i})_{\bf
10}^{m} (\Phi_{\bf 45})_{mn}{\bf 10}^n
+ y^{\prime i} ({\bf 16_i\otimes 16_i})_{\bf
120}^{mnl} (\Phi_{\bf 45})_{mn}{\bf 10}_l
\right] ~.~\,
\eeqa
We can show that the above superpotential will not contribute to
the SM fermion Yukawa coupling terms and trilinear soft terms.

For the Higgs field $\Phi_{\bf 210}$ in the ${\bf 210}$ representation,
we can write the VEV
in terms of a $16\tm 16$ matrix as follows
\beqa
\langle \Phi_{\bf 210} \rangle
~=~\f{v}{2\sqrt{6}} {\rm diag}(\underbrace{~1,~1,~1,-3}_4)~, \eeqa
which is normalized as $c=2$.
We consider the following
superpotential
\beqa
W & \supset& {1\over {M_*}}
\left[ h^i ({\bf 16_i\otimes 16_i})_{\bf
120}^{mnl} (\Phi_{\bf 210})_{mnlk}{\bf 10}^k
+ h^{\prime i} ({\bf 16_i\otimes 16_i})_{\bf
126}^{mnlpq} (\Phi_{\bf 210})_{mnlp}{\bf 10}_q
\right]
\nonumber \\ &&
+ \alpha' {T\over {M^2_*}}
\left[ y^i ({\bf 16_i\otimes 16_i})_{\bf
120}^{mnl} (\Phi_{\bf 210})_{mnlk}{\bf 10}^k
\right. \nonumber \\ && \left.
+ y^{\prime i} ({\bf 16_i\otimes 16_i})_{\bf
126}^{mnlpq} (\Phi_{\bf 210})_{mnlp}{\bf 10}_q
\right] ~.~\,
\eeqa
After $\Phi_{\bf 210}$ acquires a VEV, we obtain the additional
contributions to the Yukawa coupling terms
\beqa
W & \supset& h^{\prime i} \f{
v}{\sqrt{6}M_*}\left[{Q}_i{U}_i^cH_u - 3{L}_i{N}^c_i H_u
+ {Q}_iD_i^c H_d -3 {L}_i{E_i^c} H_d \right]~.~\,
\eeqa
The extra supersymmetry breaking trilinear soft terms are
\beqa
-{\cal L} & \supset & y^{\prime i} \f{F_T v}{\sqrt{6}M_*^2}
\left[\tl{Q}_i \tl{U}_i^c H_u -3\tl{L}_i \tl{N}^c_i H_u
+ \tl{Q}_i \tl{D}_i^c H_d -3 \tl{L}_i \tl{E}_i^c H_d \right]~.~\,
\eeqa




\subsection{ The Georgi-Glashow $SU(5)\tm U(1)'$ Model}

The $SO(10)$ gauge symmetry can be broken down to the Georgi-Glashow
$SU(5)\tm U(1)'$ gauge symmetry by giving VEVs to the Higgs
fields in the ${\bf 45}$ and ${\bf 210}$ representations.

For the Higgs field $\Phi_{\bf 45}$
in the ${\bf 45}$ representation, we can write the VEV as a ${\bf
10\tm 10}$ matrix:
\beqa
\langle \Phi_{\bf 45} \rangle ~=~\f{v}{\sqrt{10}}
{\rm
diag}(\underbrace{~1,\cdots,~1}_5,\underbrace{-1,\cdots,-1}_{5})~,~\,
\eeqa
where the normalization is $c=1$.
Using the conventions in~\cite{He:1990jw}
we obtain the non-zero components
\beqa
(\Phi_{\bf 45})_{12}=(\Phi_{\bf 45})_{34}=(\Phi_{\bf 45})_{56}
=(\Phi_{\bf 45})_{78}=(\Phi_{\bf 45})_{90}= \f{v}{\sqrt{10}}~.
\eeqa
To calculate the additional contributions to the Yukawa coupling terms
and trilinear soft terms, we consider the following
superpotential
\beqa
W & \supset& {1\over {M_*}}
\left[ h^i ({\bf 16_i\otimes 16_i})_{\bf
10}^m (\Phi_{\bf 45})_{mn}{\bf 10}^n + h^{\prime i}
({\bf 16_i\otimes 16_i})_{\bf 120}^{mnl}(\Phi_{\bf 45})_{mn}{\bf 10}_l
\right]
\nonumber \\
&&+ \alpha' {T\over {M^2_*}}
\left[ y^i ({\bf 16_i\otimes 16_i})_{\bf
10}^m (\Phi_{\bf 45})_{mn}{\bf 10}^n + y^{\prime i}
({\bf 16_i\otimes 16_i})_{\bf 120}^{mnl}(\Phi_{\bf 45})_{mn}{\bf 10}_l
\right]~.~\,
\eeqa
Note that ${\bf 120}$ is anti-symmetric representation, the
$h^{\prime i}$ and $y^{\prime i}$ terms will not contribute to
the SM fermion Yukawa coupling terms and trilinear soft terms.
After $\Phi_{\bf 210}$ acquires a VEV, we obtain the additional
contributions to the Yukawa coupling terms
\beqa
W & \supset& h^i \f{2
v}{\sqrt{10}M_*}\left[{Q}_i{U}_i^cH_u+{L}_i{N}^c_i H_u
-{Q}_iD_i^c H_d -{L}_i{E_i^c} H_d \right]~.~\,
\eeqa
The extra supersymmetry breaking trilinear soft terms are
\beqa
-{\cal L} & \supset & y^i \f{2F_T v}{\sqrt{10}M_*^2}
\left[\tl{Q}_i \tl{U}_i^c H_u+\tl{L}_i \tl{N}^c_i H_u
-\tl{Q}_i \tl{D}_i^c H_d - \tl{L}_i \tl{E}_i^c H_d \right]~.~\,
\eeqa

For the Higgs field $\Phi_{\bf 210}$
in the ${\bf 210}$ representation, we can write the VEV in the form of a
${\bf
16\tm 16}$ matrix as follows
\beqa
\langle \Phi_{\bf 210} \rangle ~=~\f{v}{2\sqrt{5}}
{\rm
diag}(\underbrace{~1,\cdots,~1}_5,\underbrace{-1,\cdots,-1}_{10},5)~,
\eeqa
where the
normalization is $c=2$. This VEV can be written in components
as follows
\beqa
(\Phi_{\bf 210})_{1234}&=&
(\Phi_{\bf 210})_{1256}=(\Phi_{\bf 210})_{1278}=
(\Phi_{\bf 210})_{1290}=
(\Phi_{\bf 210})_{3456}
=(\Phi_{\bf 210})_{3478}
\nn\\&=&
(\Phi_{\bf 210})_{3490}
=(\Phi_{\bf 210})_{5678}=(\Phi_{\bf 210})_{5690}
=(\Phi_{\bf 210})_{7890}=-\f{v}{2\sqrt{5}}~.~\,
\eeqa
We consider the following
superpotential
\beqa
W & \supset& {1\over {M_*}}
\left[ h^i ({\bf 16_i\otimes 16_i})_{\bf
120}^{mnl} (\Phi_{\bf 210})_{mnlk}{\bf 10}^k + h^{\prime i}
({\bf 16_i\otimes 16_i})_{\bf 126}^{mnlkp}(\Phi_{\bf 210})_{mnlk}{\bf
10}_p \right]
\nonumber \\
&&+ \alpha' {T\over {M^2_*}}
\left[ y^i ({\bf 16_i\otimes 16_i})_{\bf
120}^{mnl} (\Phi_{\bf 210})_{mnlk}{\bf 10}^k \right. \nonumber \\ &&
\left.
+ y^{\prime i}
({\bf 16_i\otimes 16_i})_{\bf 126}^{mnlkp}(\Phi_{\bf 210})_{mnlk}{\bf
10}_p
\right]~.~\,
\eeqa
After $\Phi_{\bf 45}$ acquires a VEV, we obtain the additional
contributions to the Yukawa coupling terms
\beqa
W & \supset& h^{\prime i} \f{
v}{\sqrt{5}M_*}\left[3 {L}_i{N}^c_i H_u
-{Q}_iU_i^c H_u \right]~.~\,
\eeqa
The extra supersymmetry breaking trilinear soft terms are
\beqa
-{\cal L} & \supset & y^{\prime i} \f{F_T v}{\sqrt{5}M_*^2}
\left[ 3 \tl{L}_i \tl{N}^c_i H_u
- \tl{Q}_i \tl{U}_i^c H_u \right]~.~\,
\eeqa

\subsection{ The Flipped $SU(5)\times U(1)_X$ Model}

The discussion for the flipped $SU(5)\times U(1)_X$ model
is similar to those for the Georgi-Glashow $SU(5)\times U(1)'$
model except that we make the following transformations
\beqa
Q_i \leftrightarrow Q_i~,~~U^c_i \leftrightarrow D_i^c~,~~
L_i \leftrightarrow L_i~,~~N^c_i \leftrightarrow E_i^c~,~~
H_d \leftrightarrow H_u~.~\,
\label{GG-FSU5}
\eeqa

Therefore, for the Higgs field in the ${\bf 45}$ representation,
we obtain the additional contributions to the SM fermion
Yukawa coupling terms and trilinear soft terms
\beqa
W & \supset& h^i \f{2
v}{\sqrt{10}M_*}\left[ {Q}_iD_i^c H_d + {L}_i{E_i^c} H_d
- {Q}_i{U}_i^cH_u - {L}_i{N}^c_i H_u \right]~,~\,
\eeqa
\beqa
-{\cal L} & \supset & y^i \f{2F_T v}{\sqrt{10}M_*^2}
\left[ \tl{Q}_i \tl{D}_i^c H_d + \tl{L}_i \tl{E}_i^c H_d
- \tl{Q}_i \tl{U}_i^c H_u -\tl{L}_i \tl{N}^c_i H_u
\right]~.~\,
\eeqa
For the Higgs field in the ${\bf 210}$ representation, we have
\beqa
W & \supset& h^{\prime i} \f{v}{\sqrt{5}M_*}\left[3 {L}_i{E}^c_i H_d
-{Q}_iD_i^c H_d \right]~,~\,
\eeqa
\beqa
-{\cal L} & \supset & y^{\prime i} \f{F_T v}{\sqrt{5}M_*^2}
\left[ 3 \tl{L}_i \tl{E}^c_i H_d
- \tl{Q}_i \tl{D}_i^c H_d \right]~.~\,
\eeqa




\subsection{The $SU(3)_C\tm SU(2)_L\tm U(1)_1\tm U(1)_{2}$ Model}

The $SO(10)$ gauge symmetry can be broken down to the
$SU(3)_C\tm SU(2)_L\tm U(1)_1\tm U(1)_{2}$ gauge symmetry
by giving VEVs to the ${\bf (24, 0)}$ component of the Higgs fields
in the ${\bf 45}$, ${\bf 54}$ and ${\bf 210}$ representations
under $SU(5)\times U(1)$,
or to the ${\bf (75, 0)}$ component of
the Higgs field in the ${\bf 210}$ representation.
In this subsection, we only study the Yukawa coupling terms and trilinear
soft
terms in the Georgi-Glashow $SU(5)\tm U(1)'$
model, ${\it i.e.}$ the gauge symmetry is
$SU(3)_C\times SU(2)_L \times U(1)_Y \times U(1)'$.
The Yukawa coupling terms and trilinear soft
terms in the flipped $SU(5)\times U(1)_X$ model
can be obtained from those in the Georgi-Glashow $SU(5)\tm U(1)'$
model by making the transformation
in Eq.~(\ref{GG-FSU5}).

First, for the Higgs field $\Phi_{\bf 45}$
in the ${\bf 45}$ representation, we can write the VEV in the form of a
${\bf
10\tm 10}$ matrix as follows
\beqa
\langle \Phi_{\bf 45} \rangle ~=~v{\sqrt{\f{3}{5}}}
{\rm diag}(~\f{1}{3},~\f{1}{3},~\f{1}{3},-\f{1}{2},
-\f{1}{2},-\f{1}{3},-\f{1}{3},-\f{1}{3},~\f{1}{2},~\f{1}{2})~,
\eeqa
which is normalized to $c=1$. It can also be written in
components as follows
\beqa
3(\Phi_{\bf 45})_{12}=3(\Phi_{\bf 45})_{34}
=3(\Phi_{\bf 45})_{56}=-2 (\Phi_{\bf 45})_{78}
=-2 (\Phi_{\bf 45})_{90}= v{\sqrt{\f{3}{5}}}~.
\eeqa
To calculate the additional contributions to the Yukawa coupling terms
and trilinear soft terms, we consider the following
superpotential
\beqa
W & \supset& {1\over {M_*}}
\left[ h^i ({\bf 16_i\otimes 16_i})_{\bf
10}^m (\Phi_{\bf 45})_{mn}{\bf 10}^n + h^{\prime i}
({\bf 16_i\otimes 16_i})_{\bf 120}^{mnl}(\Phi_{\bf 45})_{mn}{\bf 10}_l
\right]
\nonumber \\
&&+ \alpha' {T\over {M^2_*}}
\left[ y^i ({\bf 16_i\otimes 16_i})_{\bf
10}^m (\Phi_{\bf 45})_{mn}{\bf 10}^n + y^{\prime i}
({\bf 16_i\otimes 16_i})_{\bf 120}^{mnl}(\Phi_{\bf 45})_{mn}{\bf 10}_l
\right]~.~\,
\eeqa
After $\Phi_{\bf 45}$ acquires a VEV, we obtain the additional
contributions to the Yukawa coupling terms
\beqa
W & \supset& h^i \f{v}{2M_*} \sqrt{\f{3}{5}}
\left[{Q}_i{U}_i^cH_u+{L}_i{N}^c_i H_u
-{Q}_iD_i^c H_d -{L}_i{E_i^c} H_d \right]~.~\,
\eeqa
The extra supersymmetry breaking trilinear soft terms are
\beqa
-{\cal L} & \supset & y^i \f{F_T v}{2 M_*^2} \sqrt{\f{3}{5}}
\left[\tl{Q}_i \tl{U}_i^c H_u+\tl{L}_i \tl{N}^c_i H_u
-\tl{Q}_i \tl{D}_i^c H_d - \tl{L}_i \tl{E}_i^c H_d \right]~.~\,
\eeqa
The new contributions to the low-energy Yukawa coupling terms and
trilinear
soft terms are the same as the
$SU(5)\tm U(1)'$ models.

Second, for the Higgs field $\Phi_{\bf 54}$
in the ${\bf 54}$ representation, we can write the VEV in the form of a
${\bf
10\tm 10}$ matrix as follows
\beqa
\langle \Phi_{\bf 54} \rangle ~=~
v{\sqrt{\f{3}{5}}}
{\rm diag}(~\f{1}{3},~\f{1}{3},~\f{1}{3},
-\f{1}{2},-\f{1}{2},~\f{1}{3},~\f{1}{3},~\f{1}{3},-\f{1}{2},-\f{1}{2})~,
\eeqa
which is normalized to $c=1$.
We consider the following
superpotential
\beqa
W & \supset& {1\over {M_*}}
h^i ({\bf 16_i\otimes 16_i})_{\bf
10}^m (\Phi_{\bf 54})_{mn}{\bf 10}^n
+ \alpha' {T\over {M^2_*}}
y^i ({\bf 16_i\otimes 16_i})_{\bf
10}^m (\Phi_{\bf 54})_{mn}{\bf 10}^n ~.~\,
\eeqa
After $\Phi_{\bf 54}$ acquires a VEV, we obtain the additional
contributions to the Yukawa coupling terms
\beqa
W & \supset& - h^i \f{v}{2M_*} \sqrt{\f{3}{5}}
\left[{Q}_i{U}_i^cH_u+{L}_i{N}^c_i H_u
+ {Q}_iD_i^c H_d + {L}_i{E_i^c} H_d \right]~.~\,
\eeqa
The extra supersymmetry breaking trilinear soft terms are
\beqa
-{\cal L} & \supset & - y^i \f{F_T v}{2 M_*^2} \sqrt{\f{3}{5}}
\left[\tl{Q}_i \tl{U}_i^c H_u+\tl{L}_i \tl{N}^c_i H_u
+ \tl{Q}_i \tl{D}_i^c H_d + \tl{L}_i \tl{E}_i^c H_d \right]~.~\,
\eeqa

Third, we consider that the ${\bf (24, 0)}$ component
of the Higgs field $\Phi^{\bf 24}_{\bf 210}$ in the
${\bf 210}$ representation obtains a VEV.
We can write its VEV in the ${\bf
16\tm 16}$ matrix as follows
\beqa
\langle \Phi^{\bf 24}_{\bf 210} \rangle ~=~
\f{v}{\sqrt{5}} {\rm diag}(-1,-1,-1,~\f{3}{2},~\f{3}{2},
\underbrace{~\f{1}{6},\cdots,~\f{1}{6}}_6,-\f{2}{3},-\f{2}{3},-\f{2}{3},~1,~0)~,
\eeqa
which is normalized to $c=2$. In components we have
\beqa
6(\Phi^{\bf 24}_{\bf 210})_{1278}
&=& 6(\Phi^{\bf 24}_{\bf 210})_{3478}
=6(\Phi^{\bf 24}_{\bf 210})_{5678}
=6(\Phi^{\bf 24}_{\bf 210})_{1290}
\nonumber \\&=&
6(\Phi^{\bf 24}_{\bf 210})_{3490}
=6(\Phi^{\bf 24}_{\bf 210})_{5690}
=-\f{3}{2}(\Phi^{\bf 24}_{\bf 210})_{1234}
\nonumber \\&=&
=-\f{3}{2}(\Phi^{\bf 24}_{\bf 210})_{1256}
=-\f{3}{2} (\Phi^{\bf 24}_{\bf 210})_{3456}
= (\Phi^{\bf 24}_{\bf 210})_{7890}=\f{v}{\sqrt{5}} ~.
\eeqa
We consider the following
superpotential
\beqa
W & \supset& {1\over {M_*}}
\left[ h^i ({\bf 16_i\otimes 16_i})_{\bf
120}^{mnl} (\Phi^{\bf 24}_{\bf 210})_{mnlk}{\bf 10}^k
+ h^{\prime i} ({\bf 16_i\otimes 16_i})_{\bf
126}^{mnlpq} (\Phi^{\bf 24}_{\bf 210})_{mnlp}{\bf 10}_q
\right]
\nonumber \\ &&
+ \alpha' {T\over {M^2_*}}
\left[ y^i ({\bf 16_i\otimes 16_i})_{\bf
120}^{mnl} (\Phi^{\bf 24}_{\bf 210})_{mnlk}{\bf 10}^k
\right. \nonumber \\ && \left.
+ y^{\prime i} ({\bf 16_i\otimes 16_i})_{\bf
126}^{mnlpq} (\Phi^{\bf 24}_{\bf 210})_{mnlp}{\bf 10}_q
\right] ~.~\,
\eeqa
After $\Phi^{\bf 24}_{\bf 210}$ acquires a VEV, we obtain the additional
contributions to the Yukawa coupling terms
\beqa
W & \supset& h^{\prime i} \f{v}{M_*} \f{1}{6\sqrt{5}}
\left[-3 {Q}_i{U}_i^cH_u+9{L}_i{N}^c_i H_u
-5 {Q}_iD_i^c H_d +15{L}_i{E_i^c} H_d \right]~.~\,
\eeqa
The extra supersymmetry breaking trilinear soft terms are
\beqa
-{\cal L} & \supset & y^{\prime i} \f{F_T v}{ M_*^2} \f{1}{6\sqrt{5}}
\left[-3 \tl{Q}_i \tl{U}_i^c H_u+ 9 \tl{L}_i \tl{N}^c_i H_u
-5\tl{Q}_i \tl{D}_i^c H_d +15 \tl{L}_i \tl{E}_i^c H_d \right]~.~\,
\eeqa

Finally, we consider that the ${\bf (75, 0)}$ component
of the Higgs field $\Phi^{\bf 75}_{\bf 210}$ in the
${\bf 210}$ representation obtains a VEV.
We can write its VEV in the ${\bf
16\tm 16}$ matrix as follows
\beqa
\langle \Phi^{\bf 75}_{\bf 210} \rangle ~=~
\f{v}{3} {\rm
diag}(~0,~0,~0,~0,~0,\underbrace{-1,\cdots,-1}_6,~1,~1,~1,~3,~0)~,
\eeqa
which is normalized as $c=2$.
In components we have
\beqa
(\Phi^{\bf 75}_{\bf 210})_{1278}
&=& (\Phi^{\bf 75}_{\bf 210})_{3478}
=(\Phi^{\bf 75}_{\bf 210})_{5678}
=(\Phi^{\bf 75}_{\bf 210})_{1290}
\nonumber \\&=&
(\Phi^{\bf 75}_{\bf 210})_{3490}
=(\Phi^{\bf 75}_{\bf 210})_{5690}
=-(\Phi^{\bf 75}_{\bf 210})_{1234}
\nonumber \\&=&
=-(\Phi^{\bf 75}_{\bf 210})_{1256}
=- (\Phi^{\bf 75}_{\bf 210})_{3456}
=- {1\over 3} (\Phi^{\bf 75}_{\bf 210})_{7890}= -\f{v}{3} ~.
\eeqa
We consider the following
superpotential
\beqa
W & \supset& {1\over {M_*}}
\left[ h^i ({\bf 16_i\otimes 16_i})_{\bf
120}^{mnl} (\Phi^{\bf 75}_{\bf 210})_{mnlk}{\bf 10}^k
+ h^{\prime i} ({\bf 16_i\otimes 16_i})_{\bf
126}^{mnlpq} (\Phi^{\bf 75}_{\bf 210})_{mnlp}{\bf 10}_q
\right]
\nonumber \\ &&
+ \alpha' {T\over {M^2_*}}
\left[ y^i ({\bf 16_i\otimes 16_i})_{\bf
120}^{mnl} (\Phi^{\bf 75}_{\bf 210})_{mnlk}{\bf 10}^k
\right. \nonumber \\ && \left.
+ y^{\prime i} ({\bf 16_i\otimes 16_i})_{\bf
126}^{mnlpq} (\Phi^{\bf 75}_{\bf 210})_{mnlp}{\bf 10}_q
\right] ~.~\,
\eeqa
After $\Phi^{\bf 75}_{\bf 210}$ acquires a VEV, we obtain the additional
contributions to the Yukawa coupling terms
\beqa
W & \supset& h^{\prime i} \f{v}{3M_*}
\left[- {Q}_iD_i^c H_d + 3{L}_i{E_i^c} H_d \right]~.~\,
\eeqa
The extra supersymmetry breaking trilinear soft terms are
\beqa
-{\cal L} & \supset & y^{\prime i} \f{F_T v}{ 3M_*^2}
\left[-\tl{Q}_i \tl{D}_i^c H_d +3 \tl{L}_i \tl{E}_i^c H_d \right]~.~\,
\eeqa

\section{Scalar and Gaugino Mass Relations}
\label{sec-4}
In order to study the scalar and gaugino mass relations that
are invariant under one-loop renormalization group running, we need to
know the renormalization group equations (RGEs) of the supersymmetry
breaking scalar masses and gaugino masses. For simplicity, we only
consider the
one-loop RGE running since the two-loop RGE running
effects are small~\cite{Li:2010mr}.
In particular, for the first two generations, we can
neglect the contributions from the Yukawa coupling terms and trilinear
soft terms, and then the RGEs for the scalar masses
are~\cite{Martin:1993zk}
\beqa 16\pi^2\f{d m^2_{\tl{Q}_{j}}}{d
t}&=&-\f{32}{3}g_3^2M_3^2-6g_2^2M_2^2-\f{2}{15}g_1^2M_1^2+\f{1}{5}g_1^2S~,\\
16\pi^2\f{d m^2_{\tl{U}^c_{j}}}{d
t}&=&-\f{32}{3}g_3^2M_3^2-\f{32}{15}g_1^2M_1^2-\f{4}{5}g_1^2S~,\\
16\pi^2\f{d m^2_{\tl{D}^c_{j}}}{d
t}&=&-\f{32}{3}g_3^2M_3^2-\f{8}{15}g_1^2M_1^2+\f{2}{5}g_1^2S~,\\
16\pi^2\f{d m^2_{\tl{L}_{j}}}{d t}&=&
-6g_2^2M_2^2-\f{6}{5}g_1^2M_1^2-\f{3}{5}g_1^2S~,\\
16\pi^2\f{d m^2_{\tl{E}^c_{j}}}{d
t}&=&-\f{24}{5}g_1^2M_1^2+\f{6}{5}g_1^2S~, \eeqa
where $j=1,~2$, and $t={\rm ln}\mu$ and $\mu$ is the renormalization
scale.
Also, $S$ is given by
\beqa
S=Tr[Y_{\phi_i}
m^2(\phi_i)]=m_{H_u}^2-m_{H_d}^2+Tr[M_{\tl{Q}_i}^2-M_{\tl{L}_i}^2
-2M_{\tl{U}^c_i}^2+M_{\tl{D}_i^c}^2+M_{\tl{E}_i^c}^2]~.~\,
\eeqa

The one-loop RGEs for gauge couplings $g_i$ and gaugino masses
$M_i$ are
\beqa
\f{d}{dt}g_i~=~\f{1}{16\pi^2}b_ig_i^3~,~~~
\f{d}{dt}M_i~=~\f{1}{8\pi^2}b_ig_i^2M_i~,
\eeqa
where $g_1\equiv \sqrt{5} g_Y/\sqrt{3}$, and $b_1$, $b_2$ and $b_3$
are one-loop beta functions for $U(1)_Y$, $SU(2)_L$, and $SU(3)_C$,
respectively. For the supersymmetric SM, we have
\beqa
b_3=-3~,~b_2=1~,~b_1=\f{33}{5}~.
\eeqa
Therefore, we obtain
\beqa
\f{d}{dt} \[\f{MSQj}{Y_{Q_j}}\] &=&\f{d}{dt} \[\f{MSUj}{Y_{U^c_j}}\]
~=~\f{d}{dt} \[\f{MSDj}{Y_{D^c_j}}\] \nonumber \\
&=&\f{d}{dt} \[\f{MSLj}{Y_{L_j}}\] 
 ~=~\f{d}{dt} \[\f{MSEj}{Y_{E^c_j}}\]~,
\eeqa
where
\beqa
MSQj &=& 4 m^2_{\tl{Q}_{j}} + \f{32}{3b_3} M_3^2 + \f{6}{b_2} M_2^2 
+ \f{2}{15b_1} M_1^2 ~,~ \\
MSUj &=& 4 m^2_{\tl{U}^c_{j}} + \f{32}{3b_3} M_3^2 
+ \f{32}{15b_1} M_1^2 ~,~ \\
MSDj &=& 4 m^2_{\tl{D}^c_{j}} + \f{32}{3b_3} M_3^2 
+ \f{8}{15b_1} M_1^2 ~,~ \\
MSLj &=& 4 m^2_{\tl{L}_{j}} +  \f{6}{b_2} M_2^2 
+ \f{6}{5b_1} M_1^2 ~,~ \\
MSEj &=& 4 m^2_{\tl{E}^c_{j}}  + \f{24}{5b_1} M_1^2 ~.~
\eeqa
In addition, we obtain the most general scalar and gaugino mass relations
that are valid from the GUT scale to the electroweak 
scale under one-loop RGE running for the first two families
\beqa
\gamma_{Q_j} \f{MSQj}{Y_{Q_j}} + \gamma_{U^c_j} \f{MSUj}{Y_{U^c_j}}
+ \gamma_{D^c_j} \f{MSDj}{Y_{D^c_j}} + \gamma_{L_j} \f{MSLj}{Y_{L_j}}
+ \gamma_{E^c_j} \f{MSEj}{Y_{E^c_j}} = C_o~,~\,
\eeqa
where $C_o$ denotes the invariant constant under one-loop
RGE running, and
 $\gamma_{Q_j}$, $\gamma_{U^c_j}$, $\gamma_{D^c_j}$, 
$\gamma_{L_j}$, and $\gamma_{E^c_j}$ are real or complex numbers that
satisfy
\beqa
\gamma_{Q_j} + \gamma_{U^c_j} + \gamma_{D^c_j} + \gamma_{L_j}
+ \gamma_{E^c_j} ~=~ 0~.~\,
\eeqa
In this paper, we shall study the following
scalar and gaugino mass relations 
\beqa
C_o^{AB} &=& 3 m^2_{\tl{D}^c_{j}} + 2  m^2_{\tl{L}_{j}}
-4 m^2_{\tl{Q}_{j}} - m^2_{\tl{U}^c_{j}} -\[\f{16}{3b_3}M_3^2
+\f{3}{b_2} M_2^2 -\f{1}{3b_1}M_1^2 \]~,~ \\
C_o^{AC} &=& 3 m^2_{\tl{D}^c_{j}} + 2 m^2_{\tl{L}_{j}}
+  m^2_{\tl{E}^c_{j}} - 6 m^2_{\tl{Q}_{j}} 
-\[\f{8}{b_3}M_3^2 +\f{6}{b_2} M_2^2 -\f{2}{b_1}M_1^2 \] ~,~ \\
C_o^{AD} &=& 3 m^2_{\tl{D}^c_{j}}+ 2 m^2_{\tl{L}_{j}}
-3 m^2_{\tl{U}^c_{j}} -2 m^2_{\tl{E}^c_{j}}
+\[\f{3}{b_2} M_2^2 -\f{3}{b_1}M_1^2 \]~,~ \\
C_o^{BC} &=&  m^2_{\tl{U}^c_{j}} + m^2_{\tl{E}^c_{j}}
- 2 m^2_{\tl{Q}_{j}} -\[\f{8}{3b_3}M_3^2
+\f{3}{b_2} M_2^2 -\f{5}{3b_1}M_1^2 \]~,~ \\
C_o^{X} &=& m^2_{\tl{Q}_{j}} + 3 m^2_{\tl{L}_{j}}+ m^2_{\tl{E}^c_{j}}
- 2 m^2_{\tl{U}^c_{j}} - 3 m^2_{\tl{D}^c_{j}} 
-\[\f{32}{3b_3}M_3^2
-\f{6}{b_2} M_2^2 -\f{2}{3b_1}M_1^2 \] ~.~\,
\eeqa
In short, we can obtain the scalar and gaugino mass relations
that are valid from the GUT scale to the electroweak scale
at one loop. Such relations will be useful to distinguish
between the mSUGRA and GmSUGRA scenarios.

The scalar and gaugino mass relations can be simplified by the
scalar and gaugino mass relations at the GUT scale. Because the
high-dimensional operators can contribute to gauge kinetic
functions after GUT symmetry breaking,
the SM gauge couplings may not be unified at the
GUT scale. Thus, we will have two contributions
to the gaugino masses at
the GUT scale: the universal gaugino masses as in the mSUGRA,
and the non-universal gaugino masses due to the high-dimensional
operators.
In particular, for the scenarios studied in
Refs.~\cite{Anderson:1999uia, Chamoun:2001in, Chakrabortty:2008zk,
Martin:2009ad, Bhattacharya:2009wv, Feldman:2009zc, Chamoun:2009nd}
where the universal gaugino masses are assumed to be zero,
{\it i.e.}, $M_i/\alpha_i=a_i M'_{1/2}$,
we obtain the gaugino mass relation at one loop~\cite{Li:2010xr}
\beqa
\f{M_3}{a_3 \alpha_3}=\f{M_2}{a_2 \alpha_2}=\f{M_1}{a_1 \alpha_1} ~.
\eeqa
We can calculate the scalar and gaugino mass relations
in the mSUGRA and GmSUGRA scenarios, and
compare them in different cases.

\subsection{ The $SU(5)$ Model}

In the following, we consider the RGE running for the scalar masses of
the first two families
in the $SU(5)$ model with the Higgs fields in the
${\bf 24}$ and ${\bf 75}$ representations.
\begin{itemize}
\item The $SU(5)$ Model with a ${\bf 24}$ Dimensional Higgs Field

Let us consider the scalar and gaugino mass relations $C_o^{AB}$, 
$C_o^{AC}$, $C_o^{AD}$, and $C_o^{BC}$ in the  mSUGRA and GmSUGRA  scenarios
 for the first two generations.
In the mSUGRA scenario with universal gaugino masses and scalar masses,
we obtain the $C_o^{AB}$, 
$C_o^{AC}$, $C_o^{AD}$, and $C_o^{BC}$ as follows
\beqa
(C_o^{AB})^{U} &=&  -\f{116}{99}\(\f{M_1^2(\mu)}{g_1^4(\mu)}\)g_1^4(M_U) ~,~\, \\
(C_o^{AC})^{U} &=&  -\f{100}{33}\(\f{M_1^2(\mu)}{g_1^4(\mu)}\)g_1^4(M_U) ~,~\, \\
(C_o^{AD})^{U} &=&  \f{28}{11}\(\f{M_1^2(\mu)}{g_1^4(\mu)}\)g_1^4(M_U) ~,~\, \\
 (C_o^{BC})^{U} &=&  -\f{184}{99}\(\f{M_1^2(\mu)}{g_1^4(\mu)}\)g_1^4(M_U) ~.~\,
\eeqa

In the GmSUGRA scenario, we consider the scenario in
Refs.~\cite{Anderson:1999uia, Chamoun:2001in, Chakrabortty:2008zk,
Martin:2009ad, Bhattacharya:2009wv, Feldman:2009zc, Chamoun:2009nd}.
At the GUT scale we have
\beqa
\f{M_3}{2}=\f{M_2}{-3}=\f{M_1}{-1}~.
\eeqa
Thus, with Eq. (\ref{SMass-R}) for the
non-universal scalar mass relations at
the GUT scale in the GmSUGRA scenario, we obtain
the scalar and gaugino mass relations
\beqa
(C_o^{AB})^{NU} &=&  -\f{1964}{99}\(\f{M_1^2(\mu)}{g_1^4(\mu)}\)g_1^4(M_U) ~,~\, \\
(C_o^{AC})^{NU} &=&  -\f{1420}{33}\(\f{M_1^2(\mu)}{g_1^4(\mu)}\)g_1^4(M_U) ~,~\, \\
(C_o^{AD})^{NU} &=&  \f{292}{11}\(\f{M_1^2(\mu)}{g_1^4(\mu)}\)g_1^4(M_U) ~,~\, \\
 (C_o^{BC})^{NU} &=&  -\f{2296}{99}\(\f{M_1^2(\mu)}{g_1^4(\mu)}\)g_1^4(M_U) ~.~\,
\eeqa
Thus, with precise enough measurements, we may distinguish the
mSUGRA and GmSUGRA scenarios. In particular, we can consider
the ratios of these one-loop RGE invariant constants and then distinguish
the mSUGRA and GmSUGRA scenarios, for example,
$(C_o^{AC})^{U}/(C_o^{AB})^{U} = 2.586$,
while $(C_o^{AC})^{NU}/(C_o^{AB})^{NU}=2.169 $.
Similarly, we can discuss the other scalar and gaugino mass
relations for the first two generations in mSUGRA and GmSUGRA.

\item The $SU(5)$ Model with a ${\bf 75}$ Dimensional Higgs Field

In the mSUGRA scenario with universal gaugino and scalar masses, we
obtain the one-loop RGE invariant constant $C_o^X$ at the GUT scale
\beqa
(C_o^X)^U &=& \f{956}{99}\(\f{M_3^2(\mu)}{g_3^4(\mu)}\)g_3^4(M_U)~.~\,
\eeqa

In the GmSUGRA scenario with
non-universal gaugino and scalar masses, we consider
the non-universal gaugino mass ratios in
Refs.~\cite{Anderson:1999uia, Chamoun:2001in, Chakrabortty:2008zk,
Martin:2009ad, Bhattacharya:2009wv, Feldman:2009zc, Chamoun:2009nd}
\beqa
\f{M_3}{1}=\f{M_2}{3}=\f{M_1}{-5}~.
\eeqa
With the non-universal scalar masses in Eq.~(\ref{SUV-75SM}), we obtain
\beqa
(C_o^X)^{NU} &=& \f{5948}{99}\(\f{M_3^2(\mu)}{g_3^4(\mu)}\)g_3^4(M_U)~.~\,
\eeqa
Assuming that there are no threshold corrections from the
electroweak scale to the GUT scale, we can calculate the
gauge couplings at the GUT scale and
check these scalar and gaugino mass relations
if we know the low energy sparticle spectrum.

\end{itemize}
\subsection{ The Pati-Salam Model from $SO(10)$ }

We consider the following $SO(10)$ gauge symmetry breaking chain
\beqa
SO(10)&\ra& SU(4)_C\tm SU(2)_L\tm SU(2)_R\ra SU(3)_C\tm SU(2)_L\tm
SU(2)_R\tm U(1)_{B-L}\nn\\&\ra& SU(3)_C\tm SU(2)_L\tm U(1)_Y~. \eeqa
Other symmetry breaking chains can be discussed similarly.

Let us explain our convention. We denote the gauge couplings for
the $SU(2)_L$, $SU(2)_R$, $U(1)_{B-L}$, and $SU(4)_C$
gauge symmetries as $g_{2L}$,
$g_{2R}$, $\tl{g}_{B-L}$ (or traditional
$g_{B-L}$), and $g_4$, respectively.
We denote the gaugino masses for the
$SU(2)_L$, $SU(2)_R$, $U(1)_{B-L}$, and $SU(4)_C$ gauge symmetries as
$M_{2L}$, $M_{2R}$, $M_{B-L}$ and $M_4$, respectively.
We denote the one-loop beta functions for the
$SU(2)_L$, $SU(2)_R$, $U(1)_{B-L}$, and $SU(4)_C$ gauge symmetries as
$b_{2L}$, $b_{2R}$, $\tl{b}_{B-L}$ and $b_4$, respectively.
In addition, we denote the universal supersymmetry breaking
scale as $M_S$, the $SU(2)_R\times U(1)_{B-L}$ gauge
symmetry breaking scale as $M_{LR}$, and the $SU(4)_C$ gauge
symmetry breaking scale as $M_{PS}$. Also, we denote
the $U(1)_{B-L}$ charge for the particle $\phi_i$ as $Y^{B-L}_{\phi_i}$.

The generator $U(1)_{B-L}$ in $SU(4)_C$ is
\beqa
\tl{g}^{B-L}T^{B-L}~=~
g_{B-L} {\rm diag} \left(\f{1}{6}, \f{1}{6}, \f{1}{6}, -\f{1}{2}
\right)~.~\,
\eeqa
So we can obtain the normalization of $g_{B-L}$ into $SU(4)_C$ \beqa
g_{B-L}=\sqrt{\f{3}{2}}\tl{g}^{B-L}~. \eeqa

Neglecting the Yukawa coupling terms and trilinear soft terms,
we obtain the RGEs for the scalar masses of the first two
generations in the Pati-Salam model
\beqa 16\pi^2\f{d m^2_{\tl{F}_j^L}}{dt}
&=&4\pi\f{d}{dt}\[-\f{15}{b_4}M_4^2-\f{6}{b_{2L}}M_{2L}^2\]~,\\16\pi^2\f{d
m^2_{\tl{F}_j^{Rc}}}{dt}&=&4\pi\f{d}{dt}\[-\f{15}{b_4}M_4^2-\f{6}{b_{2R}}M_{2R}^2\]~,
\eeqa which gives \beqa
\f{d}{dt}\[4m^2_{\tl{F}_j^L}+\f{15}{b_4}M_4^2+\f{6}{b_{2L}}M_{2L}^2\]&=&0~,\\
\f{d}{dt}\[4m^2_{\tl{F}_j^{Rc}}+\f{15}{b_4}M_4^2+\f{6}{b_{2R}}M_{2R}^2\]&=&0~.
\eeqa
The RGEs of the scalar masses for the
first two generations in the
$SU(3)_C\tm SU(2)_L\tm SU(2)_R\tm U(1)_{B-L}$ model are
\beqa 16\pi^2\f{d m^2_{\tl{Q}_{j}}}{d
t}&=&-\f{32}{3}g_3^2M_3^2-6g_{2L}^2M_{2L}^2-\f{1}{3}\tl{g}_{B-L}^2M_{B-L}^2+\f{1}{2}\tl{g}_{B-L}^2S^\pr~,\\
16\pi^2\f{d m^2_{\tl{U}_j^c,\tl{D}_{j}^c}}{d
t}&=&-\f{32}{3}g_3^2M_3^2-6g_{2R}^2M_{2R}^2-\f{1}{3}\tl{g}_{B-L}^2M_{B-L}^2-\f{1}{2}\tl{g}_{B-L}^2S^\pr~,\\
16\pi^2\f{d m^2_{\tl{L}_{j}}}{d t}&=&
-6g_{2L}^2M_{2L}^2-3\tl{g}_{B-L}M_{B-L}^2-\f{3}{2}\tl{g}_{B-L}^2S^\pr~,\\
16\pi^2\f{d m^2_{\tl{E}^c_{j}}}{d
t}&=&-6g_{2R}^2M_{2R}^2-3\tl{g}_{B-L}M_{B-L}^2+\f{3}{2}\tl{g}_{B-L}^2S^\pr~,\eeqa
where
\beqa
S^\pr=Tr[Y_{\phi_i}^{B-L}m^2(\phi_i)]~.
\eeqa

We consider the following linear combination of the squared scalar masses
\beqa
&&16\pi^2\f{d}{dt}\(m_{\tl{U}_j^c}^2+m_{\tl{E}_j^c}^2-2m_{\tl{Q}_j}^2\)
\nonumber \\
=&&4\pi^2\f{d}{dt}\[\f{32}{3b_3}M_3^2+\f{12}{b_{2L}}M_{2L}^2-\f{20}{3{b}_{1}}
M_{1}^2\]~~~{\rm for }~~M_{S}<\mu<M_{LR} \nonumber \\
=&&4\pi^2\f{d}{dt}\[\f{32}{3b_3}M_3^2
+\f{12}{b_{2L}}M_{2L}^2-\f{12}{b_{2R}}M_{2R}^2-\f{8}{3\tl{b}_{B-L}}
M_{B-L}^2\] \nonumber \\ &&
~~~{\rm for }~~M_{LR}<\mu<M_{PS} \nonumber \\
=&&4\pi^2\f{d}{dt}\[\f{12}{b_{2L}}M_{2L}^2-\f{12}{b_{2R}}M_{2R}^2\]
~~~{\rm for }~~M_{PS}<\mu<M_{U}~.~\,
\eeqa
From this, we obtain the scalar and gaugino mass relations which are
exact
from the GUT scale to the supersymmetry breaking scale at one loop
\beqa
&&4\(m_{\tl{U}_j^c}^2+m_{\tl{E}_j^c}^2-2m_{\tl{Q}_j}^2\)
-\f{32}{3b_3}M_3^2-\f{12}{b_{2L}}M_{2L}^2+\f{20}{3b_1}M_1^2=C_o^1~,\\
&&4\(m_{\tl{U}_j^c}^2+m_{\tl{E}_j^c}^2-2m_{\tl{Q}_j}^2\)-\f{32}{3b_3}M_3^2
-\f{12}{b_{2L}}M_{2L}^2+\f{12}{b_{2R}}M_{2R}^2
\nonumber \\ &&
+\f{8}{3\tl{b}_{B-L}}M_{B-L}^2 =C_o^2~,\\
&&4\(m_{\tl{U}_j^c}^2+m_{\tl{E}_j^c}^2-2m_{\tl{Q}_j}^2\)-\f{12}{b_{2L}}M_{2L}^2+\f{12}{b_{2R}}M_{2R}^2=C_o^3~.
\eeqa
The differences between the constants $C_o^1$ and $C_o^2$ and
between the constants $C_o^2$ and $C_o^3$ are the
threshold contributions from the extra particles due to gauge
symmetry breaking. Thus, the three constants can be determined by
matching the threshold contributions at the symmetry breaking scales.
The difference between $C_o^2$ and $C_o^3$ is
\beqa
C_o^2-C_o^3
&=&-\(\f{32}{3b_3}-\f{8}{3\tl{b}_{B-L}}\)\f{M_3^2(\mu)}{g_3^2({\mu})}g_3^2(M_{PS})~,
\eeqa
which can be determined at the scale $M_{PS}$. At this $SU(3)_C\times
U(1)_{B-L}$
unification scale, we have
\beqa
\f{M_3}{g_3^2}=\f{M_{B-L}}{\tl{g}_{B-L}^2}=\f{M_4}{g^2_4}~.~\,
\eeqa

For mSUGRA with universal gaugino and scalar masses, we have
\beqa
\f{M_3}{g_3^2}=\f{M_{B-L}}{\tl{g}_{B-L}^2}=\f{M_{2L}}{g_{2L}^2}=\f{M_{2R}}{g_{2R}^2}=\f{M_4}{g_4^2}~.
\eeqa
Thus, we can get the scalar and gaugino mass relations in supersymmetric
Standard
Models
\beqa
&&4\(m_{\tl{U}_j^c}^2+m_{\tl{E}_j^c}^2-2m_{\tl{Q}_j}^2\)
-\f{32}{3b_3}M_3^2-\f{12}{b_{2L}}M_2^2+\f{20}{3b_1}M_1^2\nn\\
&=&\(\f{8}{3\tl{b}_{B-L}}-\f{32}{3b_3}\)\f{M_3^2(\mu)}{g_3^4({\mu})}g_3^4(M_{PS})
+\f{20}{3b_1}\f{M_1^2(\mu)}{g_1^4({\mu})}g_1^4(M_{LR})\nn\\
&-&\(\f{12}{b_{2R}}g_{2R}^4(M_{LR})+
\f{8}{3\tl{b}_{B-L}}g_{B-L}^4(M_{LR})\)\f{M_3^2(\mu)}{g_3^4({\mu})}~.\eeqa
If we know the low energy sparticle spectrum at the LHC and ILC
and $g_1^2(M_{LR})$ from the RGE running, we can get the coefficients
\beqa
c&=&\(\f{8}{3\tl{b}_{B-L}}-\f{32}{b_3}\)g_3^4(M_{PS})-\(\f{12}{b_{2R}}g_{2R}^4(M_{LR})+
\f{8}{3\tl{b}_{B-L}}g_{B-L}^4(M_{LR})\)~~, \eeqa by fitting the
experimental data.

For GmSUGRA with
non-universal gaugino and scalar masses, we consider the
Higgs field in the ${\bf 210}$ representation whose singlet component
$({\bf 1,1,1})$ acquires a VEV. To give mass to the gluino, we require
that
the universal gaugino mass be non-zero.
From Eq.~(\ref{PSsm-210}), we obtain
\beqa
m_{\tl{E}_j^c}^2+m_{\tl{U}_j^c}^2-2m_{\tl{Q}_j}^2=-\sqrt{2}\beta'_{\bf
210}\f{v
|F_T|^2}{M_*^3}~.
\eeqa
Thus, the constant combination in the supersymmetric Standard Model is
\beqa
&&4\(m_{\tl{U}_j^c}^2+m_{\tl{E}_j^c}^2-2m_{\tl{Q}_j}^2\)-\f{32}{3b_3}M_3^2-\f{12}{b_{2L}}M_{2L}^2
+\f{20}{3b_1}M_1^2\nn\\
&=&-\sqrt{2} \beta'_{\bf 210}\f{v
|F_T|^2}{M_*^3}+\(\f{8}{3\tl{b}_{B-L}}-\f{32}{3b_3}\)\f{M_3^2(\mu)}{g_3^4({\mu})}g_3^4(M_{PS})+
\f{20}{3b_1}\f{M_1^2(\mu)}{g_1^4({\mu})}g_1^4(M_{LR})\nn\\&-&\(\f{12}{b_{2R}}g_{2R}^4(M_{LR})
\f{M_{2R}^2(\mu)}{g_{2R}^4({\mu})}+
\f{8}{3\tl{b}_{B-L}}g_{B-L}^4(M_{LR})\f{M_3^2(\mu)}{g_3^4({\mu})}\)~.
\eeqa
Therefore, the scalar and gaugino mass relations
in mSUGRA are different from those in GmSUGRA.
Moreover, we can break the $SO(10)$ gauge symmetry
down to the $SU(3)_C\tm SU(2)_L\tm SU(2)_R\tm
U(1)_{B-L}$ gauge symmetry by giving VEVs to
the ({\bf 15,1,1}) components of the Higgs field
in the ${\bf 45}$ and ${\bf 210}$ representations
under $SU(4)_C\times SU(2)_L \times SU(2)_R$.
Because the discussions are similar, we will
not present them here.

\subsection{ The Flipped $SU(5)\times U(1)_X$ Model from $SO(10)$ }

In the flipped $SU(5)\times U(1)_X$ model with $SO(10)$ origin,
we have two-step gauge coupling unification: the $SU(3)_C\times SU(2)_L$
gauge symmetry is unified at the scale $M_{23}$, and then
the $SU(5)\times U(1)_X$ gauge symmetry is unified at the scale $M_U$.
At the $M_{23}$ scale, we have the following gauge coupling relation
\beqa
\f{1}{{g_1}^2}=\f{24}{25} \f{1}{g_{1X}^2}+\f{1}{25} \f{1}{g_{5}^2}~,
\eeqa
where $g_{1X}$ and $g_5$ are the gauge couplings for the $U(1)_X$
and $SU(5)$ gauge symmetries, respectively.

Our conventions in this section are as follows. We denote the gaugino
masses
for the $U(1)_X$ and $SU(5)$ gauge symmetries as
$M_{1X}$ and $M_5$, respectively.
We denote the one-loop beta functions for
the $U(1)_X$ and $SU(5)$ gauge symmetries as $b_{1X}$ and $b_5$,
respectively.
Also, we denote
the $U(1)_{X}$ charge for the particle $\phi_i$ as $Y^{X}_{\phi_i}$.
With this notation, the RGEs for the scalar masses of the first two
generations
from the scale $M_{23}$ to $M_U$ are
\beqa 16\pi^2\f{d m^2_{\tl{F}_{j}}}{d
t}&=&-\f{144}{5}g_5^2M_5^2-\f{1}{5}g_{1X}^2M_{1X}^2+\f{1}{20}g_{1X}^2\tilde{S}~,\\
16\pi^2\f{d m^2_{\tl{\overline{f}}_{j}}}{d
t}&=&-\f{96}{5}g_5^2M_5^2-\f{9}{5}g_{1X}^2M_{1X}^2-\f{3}{20}g_{1X}^2\tilde{S}~,\\
16\pi^2\f{d m^2_{\tl{\overline{l}}_{j}}}{d
t}&=&-5g_{1X}^2M_{1X}^2+\f{1}{4}g_{1X}^2\tilde{S}~,\eeqa
where $\tilde{S}$ is
\beqa
\tilde{S}=Tr[Y^X_{\phi_i}m^2(\phi_i)]~. \eeqa

We consider the following scalar and gaugino mass relation
\beqa &&16\pi^2\f{d}{dt}\(
m_{\tl{E}_j^c}^2+m_{\tl{U}_j^c}^2-2m_{\tl{Q}_j}^2\)\\
&=&4\pi^2\f{d}{dt}\[\f{32}{3b_3}M_3^2+\f{12}{b_2}M_2^2-\f{20}{3b_1}{M}_{1}^2\]~~~
{\rm for}~~M_S<\mu<M_{23} \\
&=&4\pi^2\f{d}{dt}\[\f{192}{5b_5}M_5^2-\f{32}{5b_{1X}}M_{1X}^2\]~~~
{\rm for}~~M_{23}<\mu<M_{U}~.~\,
\eeqa

In the mSUGRA with universal gaugino and scalar masses,
we have at the scale $M_U$
\beqa
&&4\(m_{\tl{U}_j^c}^2+m_{\tl{E}_j^c}^2-2m_{\tl{Q}_j}^2\)-\(\f{192}{5b_5}M_5^2
-\f{32}{5b_{1X}}M_{1X}^2\)\\
=&&-\(\f{192}{5b_5}-\f{32}{5b_{1X}}\)\f{M_3^2(\mu)}{g_3^4({\mu})}g_3^4(M_U)~.
\eeqa
So we get the scalar and gaugino mass relation
in supersymmetric Standard Models
\beqa
&&4\(m_{\tl{U}_j^c}^2+m_{\tl{E}_j^c}^2-2m_{\tl{Q}_j}^2\)-\f{32}{3b_3}M_3^2-\f{12}{b_{2}}M_2^2+\f{20}{3b_1}M_1^2\nn\\
&=&-\(\f{32}{3b_3}+\f{12}{b_2}-\f{192}{5b_5}+\f{32}{5b_{1X}}\)\f{M_3^2(\mu)}{g_3^4({\mu})}g_3^4(M_{23})+\f{20}{3b_1}\f{M_1^2(\mu)}{g_1^4({\mu})}g_1^4(M_{23})
\nonumber \\
&&
-\(\f{192}{5b_5}-\f{32}{5b_{1X}}\)\f{M_3^2(\mu)}{g_3^4({\mu})}g_3^4(M_U)~.~\,
\eeqa

In GmSUGRA with
non-universal gaugino and scalar masses, we consider the
Higgs field in the ${\bf 210}$ representation whose singlet component
$({\bf 1,0})$ acquires a VEV. For non-universal gaugino masses, we
consider the
mass ratios in
Refs.~\cite{Anderson:1999uia, Chamoun:2001in, Chakrabortty:2008zk,
Martin:2009ad, Bhattacharya:2009wv, Feldman:2009zc, Chamoun:2009nd}
\beqa \f{M_5}{-1}=\f{M_{1X}}{4}~. \eeqa

With Eq.~(\ref{FSUV-210}), we obtain at the scale $M_{U}$
\beqa &&4\(m_{\tl{U}_j^c}^2+m_{\tl{E}_j^c}^2-2m_{\tl{Q}_j}^2\)
-\(\f{192}{5b_5}M_5^2-\f{32}{5b_{1X}}M_{1X}^2\)\\
=&& \f{16}{\sqrt{5}}
\beta'_{\bf 210}\f{v|F_T|^2}{M^3}
-\(\f{192}{5b_5}-\f{512}{5b_{1X}}\)\f{M_3^2(\mu)}{g_3^4({\mu})}g_3^4(M_U)~.
\eeqa
Thus, we obtain the scalar and gaugino mass relation in supersymmetric
Standard
Models \beqa
&&4\(m_{\tl{U}_j^c}^2+m_{\tl{E}_j^c}^2-2m_{\tl{Q}_j}^2\)-\f{32}{3b_3}M_3^2-\f{12}{b_{2}}M_2^2+\f{20}{3b_1}M_1^2\nn\\
&=&\f{16}{\sqrt{5}}
\beta'_{\bf 210}\f{v|F_T|^2}{M^3}
-\(\f{192}{5b_5}-\f{512}{5b_{1X}}\)\f{M_3^2(\mu)}{g_3^4({\mu})}g_3^4(M_U)\nn\\
&&-\(\f{32}{3b_3}+\f{12}{b_2}-\f{192}{5b_5}+\f{512}{5b_{1X}}\)\f{M_3^2(\mu)}{g_3^4({\mu})}g_3^4(M_{23})
+\f{20}{3b_1}\f{M_1^2(\mu)}{g_1^4({\mu})}g_1^4(M_{23})~.\eeqa
Therefore,
the dependence on ${M_3^2(\mu)}/{g_3^4({\mu})}$ in mSUGRA
is indeed different from that in GmSUGRA.
Other gauge symmetry breaking chains can be
discussed similarly.

\section{Conclusions}
\label{sec-5}

In the GmSUGRA scenario with the
high-dimensional operators containing the GUT Higgs fields,
we systematically studied the supersymmetry breaking scalar masses,
SM fermion Yukawa coupling terms,
and trilinear soft terms in the $SU(5)$ model with
GUT Higgs fields in the ${\bf 24}$ and ${\bf 75}$
representations, and in the $SO(10)$ model where
the gauge symmetry is broken down to the Pati-Salam
$SU(4)_C\times SU(2)_L \times SU(2)_R$ gauge symmetry,
$SU(3)_C\times SU(2)_L \times SU(2)_R \times U(1)_{B-L}$
gauge symmetry, George-Glashow $SU(5)\times U(1)'$
gauge symmetry, flipped $SU(5)\times U(1)_X$ gauge symmetry,
and $SU(3)_C\times SU(2)_L \times U(1)_1 \times U(1)_2$
gauge symmetry. In addition, we considered the scalar and
gaugino mass relations, which can be preserved from
the GUT scale to the electroweak scale
under one-loop RGE running, in the $SU(5)$
model, the Pati-Salam model and the flipped $SU(5)\times U(1)_X$
model arising from the $SO(10)$ model.
With such relations, we may distinguish the
mSUGRA and GmSUGRA scenarios if we can measure
the supersymmetric
particle spectrum at the LHC and ILC.
Thus, it provides us with
another important window of opportunity at the Planck scale.

Note added: after our paper was submitted, we noticed the 
paper~\cite{Carena:2010gr}, which also studies the RGE
invariants in the supersymmetric Standard Models.

\begin{acknowledgments}


This research was supported in part by the Australian 
Research Council under
project DP0877916 (CB and FW), 
by the DOE grant DE-FG03-95-Er-40917 (TL and DVN),
by the Natural Science Foundation of China
under grant No. 10821504 (TL),
and by the Mitchell-Heep Chair in High Energy Physics (TL).
CB also thanks H. Baer and X. Tata for fruitful discussions, and the
Galileo Galilei Institute for Theoretical Physics for the hospitality as
well as the INFN for partial support during the completion of this work.

\end{acknowledgments}


\begin{thebibliography}{99}
\vspace{-1mm}











\bibitem{Ellis:1990zq}
J.~R.~Ellis, S.~Kelley and D.~V.~Nanopoulos,
Phys.\ Lett.\ B {\bf 249}, 441 (1990);
Phys.\ Lett.\ B {\bf 260}, 131 (1991);
U.~Amaldi, W.~de Boer and H.~Furstenau,
coupling
Phys.\ Lett.\ B {\bf 260}, 447 (1991);
P.~Langacker and M.~X.~Luo,
$\rho_{0}$,
Phys.\ Rev.\ D {\bf 44}, 817 (1991).


\bibitem{Georgi:1974sy}
H.~Georgi and S.~L.~Glashow,
Phys.\ Rev.\ Lett.\ {\bf 32}, 438 (1974).


\bibitem{so10} H. Georgi, ``Particles And Fields: Williamsburg 1974.
AIP Conference Proceedings No. 23'', Editor C.~E.~Carlson;
H.~Fritzsch and P.~Minkowski,
Annals Phys.\ {\bf 93}, 193 (1975);
H.~Georgi and D.~V.~Nanopoulos,
Nucl.\ Phys.\ B {\bf 155}, 52 (1979).

\bibitem{mSUGRA}
A.~H.~Chamseddine, R.~L.~Arnowitt and P.~Nath,
Phys.\ Rev.\ Lett.\ {\bf 49}, 970 (1982);
H.~P.~Nilles,
Phys.\ Lett.\ B {\bf 115}, 193 (1982);
L.~E.~Ibanez,
Phys.\ Lett.\ B {\bf 118}, 73 (1982);
R.~Barbieri, S.~Ferrara and C.~A.~Savoy,
Phys.\ Lett.\ B {\bf 119}, 343 (1982);
H.~P.~Nilles, M.~Srednicki and D.~Wyler,
Phys.\ Lett.\ B {\bf 120}, 346 (1983);
J.~R.~Ellis, D.~V.~Nanopoulos and K.~Tamvakis,
Phys.\ Lett.\ B {\bf 121}, 123 (1983);
J.~R.~Ellis, J.~S.~Hagelin, D.~V.~Nanopoulos and K.~Tamvakis,
Supergravity,''
Phys.\ Lett.\ B {\bf 125}, 275 (1983);
L.~J.~Hall, J.~D.~Lykken and S.~Weinberg,
Phys.\ Rev.\ D {\bf 27}, 2359 (1983).




\bibitem{Ellis:1984bm}
J.~R.~Ellis, C.~Kounnas and D.~V.~Nanopoulos,
Nucl.\ Phys.\ B {\bf 247}, 373 (1984).



\bibitem{gaugemediation}
M.~Dine, W.~Fischler and M.~Srednicki,
Nucl.\ Phys.\ B {\bf 189}, 575 (1981); S.~Dimopoulos and S.~Raby,
Nucl.\ Phys.\ B {\bf 192}, 353 (1981); M.~Dine and W.~Fischler,
Phys.\ Lett.\ B {\bf 110}, 227 (1982);
M. Dine and A. E. Nelson,
Phys. Rev. {\bf D48}, 1277 (1993);
M. Dine,
A. E. Nelson and Y. Shirman, Phys. Rev. {\bf D51}, 1362 (1995);
M. Dine, A. E. Nelson, Y. Nir and Y.
Shirman, Phys. Rev. {\bf D53}, 2658 (1996);
for a review, see G. F. Giudice and R. Rattazzi, Phys.
Rept. {\bf 322}, 419 (1999).


\bibitem{anomalymediation}
L.~Randall and R.~Sundrum,
Nucl.\ Phys.\ B {\bf 557}, 79 (1999);
G.~F.~Giudice, M.~A.~Luty, H.~Murayama and R.~Rattazzi,
JHEP {\bf 9812}, 027 (1998).



\bibitem{UVI-AMSB}
I.~Jack and D.~R.~T.~Jones,
breaking,''
Phys.\ Lett.\ B {\bf 482}, 167 (2000);
N.~Arkani-Hamed, D.~E.~Kaplan, H.~Murayama and Y.~Nomura,
JHEP {\bf 0102}, 041 (2001);
R.~Kitano, G.~D.~Kribs and H.~Murayama,
Phys.\ Rev.\ D {\bf 70}, 035001 (2004).


\bibitem{D-AMSB}
A. Pomarol and R. Rattazzi, JHEP {\bf 9905}, 013 (1999);
R. Rattazzi, A. Strumia and J. D. Wells, Nucl. Phys. B {\bf 576}, 3;
N. Okada, Phys. Rev. {\bf D65}, 115009 (2002).



\bibitem{Ellis:1985jn}
J.~R.~Ellis, K.~Enqvist, D.~V.~Nanopoulos and K.~Tamvakis,
Phys.\ Lett.\ B {\bf 155}, 381 (1985).



\bibitem{Choi:2007ka}
K.~Choi and H.~P.~Nilles,
JHEP {\bf 0704}, 006 (2007).



\bibitem{Cho:2007qv}
W.~S.~Cho, K.~Choi, Y.~G.~Kim and C.~B.~Park,
Phys.\ Rev.\ Lett.\ {\bf 100}, 171801 (2008);
M.~M.~Nojiri, Y.~Shimizu, S.~Okada and K.~Kawagoe,
JHEP {\bf 0806}, 035 (2008).


\bibitem{Barger:1999tn}
V.~D.~Barger, T.~Han, T.~Li and T.~Plehn,
Phys.\ Lett.\ B {\bf 475}, 342 (2000).




\bibitem{Hill:1983xh}
C.~T.~Hill,
Phys.\ Lett.\ B {\bf 135}, 47 (1984).


\bibitem{Shafi:1983gz}
Q.~Shafi and C.~Wetterich,
Phys.\ Rev.\ Lett.\ {\bf 52}, 875 (1984).







\bibitem{Drees:1985bx}
M.~Drees,
Nonminimal
Phys.\ Lett.\ B {\bf 158}, 409 (1985).

\bibitem{Anderson:1999uia}
G.~Anderson, H.~Baer, C.~h.~Chen and X.~Tata,
with
Phys.\ Rev.\ D {\bf 61}, 095005 (2000).


\bibitem{Chamoun:2001in}
N.~Chamoun, C.~S.~Huang, C.~Liu and X.~H.~Wu,
Nucl.\ Phys.\ B {\bf 624}, 81 (2002).


\bibitem{Chakrabortty:2008zk}
J.~Chakrabortty and A.~Raychaudhuri,
Phys.\ Lett.\ B {\bf 673}, 57 (2009).


\bibitem{Martin:2009ad}
S.~P.~Martin,
Phys.\ Rev.\ D {\bf 79}, 095019 (2009).

\bibitem{Bhattacharya:2009wv}
S.~Bhattacharya and J.~Chakrabortty,
Unified
Phys.\ Rev.\ D {\bf 81}, 015007 (2010).

\bibitem{Feldman:2009zc}
D.~Feldman, Z.~Liu and P.~Nath,
Signatures,''
Phys.\ Rev.\ D {\bf 80}, 015007 (2009).


\bibitem{Chamoun:2009nd}
N.~Chamoun, C.~S.~Huang, C.~Liu and X.~H.~Wu,
arXiv:0909.2374 [hep-ph].




\bibitem{Vafa:1996xn}
C.~Vafa,
Nucl.\ Phys.\ B {\bf 469}, 403 (1996).


\bibitem{Donagi:2008ca}
R.~Donagi and M.~Wijnholt,
arXiv:0802.2969 [hep-th].

\bibitem{Beasley:2008dc}
C.~Beasley, J.~J.~Heckman and C.~Vafa,
JHEP {\bf 0901}, 058 (2009).

\bibitem{Beasley:2008kw}
C.~Beasley, J.~J.~Heckman and C.~Vafa,
Predictions,''
JHEP {\bf 0901}, 059 (2009).


\bibitem{Donagi:2008kj}
R.~Donagi and M.~Wijnholt,
arXiv:0808.2223 [hep-th].




\bibitem{Font:2008id}
A.~Font and L.~E.~Ibanez,
JHEP {\bf 0902}, 016 (2009).





\bibitem{Jiang:2009zza}
J.~Jiang, T.~Li, D.~V.~Nanopoulos and D.~Xie,
Phys.\ Lett.\ B {\bf 677}, 322 (2009).


\bibitem{Blumenhagen:2008aw}
R.~Blumenhagen,
Phys.\ Rev.\ Lett.\ {\bf 102}, 071601 (2009).









\bibitem{Jiang:2009za}
J.~Jiang, T.~Li, D.~V.~Nanopoulos and D.~Xie,
Nucl.\ Phys.\ B {\bf 830}, 195 (2010).






\bibitem{Li:2009cy}
T.~Li,
arXiv:0905.4563 [hep-th].



\bibitem{Leontaris:2009wi}
G.~K.~Leontaris and N.~D.~Tracas,
scale in
arXiv:0912.1557 [hep-ph].


\bibitem{Li:2010mr}
T.~Li, J.~A.~Maxin and D.~V.~Nanopoulos,
arXiv:1002.1031 [hep-ph].




\bibitem{Li:2010xr}
T.~Li and D.~V.~Nanopoulos,
arXiv:1002.4183 [hep-ph].

\bibitem{Li:2010hi}
T.~Li and D.~V.~Nanopoulos,
Unified
arXiv:1005.3798 [hep-ph].






\bibitem{smbarr} S. M. Barr,
Phys.\ Lett.\ B {\bf 112}, 219 (1982).


\bibitem{dimitri}
J.~P.~Derendinger, J.~E.~Kim and D.~V.~Nanopoulos,
Phys.\ Lett.\ B {\bf 139}, 170 (1984).

\bibitem{AEHN-0}
I.~Antoniadis, J.~R.~Ellis, J.~S.~Hagelin and D.~V.~Nanopoulos,
Phys.\ Lett.\ B {\bf 194}, 231 (1987).



\bibitem{Slansky:1981yr}
R.~Slansky,
Phys.\ Rept.\ {\bf 79}, 1 (1981).


\bibitem{He:1990jw}
X.~G.~He and S.~Meljanac,
Phys.\ Rev.\ D {\bf 41}, 1620 (1990).


\bibitem{Martin:1993zk}
S.~P.~Martin and M.~T.~Vaughn,
Breaking
Phys.\ Rev.\ D {\bf 50}, 2282 (1994)
[Erratum-ibid.\ D {\bf 78}, 039903 (2008)], and references therein.


\bibitem{Carena:2010gr}
  M.~Carena, P.~Draper, N.~R.~Shah and C.~E.~M.~Wagner,
  arXiv:1006.4363 [hep-ph].







\end{thebibliography}
\end{document}